\documentclass[epj]{svjour}

\usepackage[english]{babel}
\usepackage{hyperref}

\usepackage[pdftex]{graphicx}
\usepackage{color}

\usepackage{amsmath}
\usepackage{amssymb}
\usepackage{mathtools}
\usepackage{braket}

\usepackage[sort&compress]{natbib}
\bibpunct{[}{]}{,}{n}{}{}

\newcommand{\E}{\mathrm{e}}
\newcommand{\un}[1]{\ensuremath{\,\mathrm{#1}}}
\newcommand{\tx}[1]{\ensuremath{\text{#1}}}
\newcommand{\fig}[1]{figure~\ref{fig:#1}}

\newcommand{\eq}[1]{(\ref{eq:#1})}
\newcommand{\lr}[1]{\ensuremath{\left( #1 \right)}}
\renewcommand{\d}{\mathrm{d}}

\renewcommand{\Im}[1]{\ensuremath{\mathrm{Im} \left(#1\right)}}
\newcommand{\I}{\mathrm{i}}
\newcommand{\Gm}{\Gamma}
\newcommand{\om}{\omega}
\newcommand{\mc}{\mathcal}
\newcommand{\Abs}[1]{\ensuremath{\left| #1 \right|}}
\newcommand{\Tr}[1]{\ensuremath{\mathrm{Tr}\left(#1\right)}}

\newcommand{\pd}{\partial}
\newcommand{\ovs}{\overset}

\begin{document}

\title{Edge magnetotransport in graphene: A combined analytical and numerical study}

\author{Thomas Stegmann\inst{1,2} \fnmsep \thanks{\email{stegmann@fis.unam.mx}} \and Axel Lorke\inst{2}}

\authorrunning{T. Stegmann et al.}
\titlerunning{Edge magnetotransport in graphene}

\institute{Instituto de Ciencias F\'{i}sicas, Universidad Nacional Aut\'{o}noma de M\'{e}xico, Cuernavaca, M\'{e}xico \and 
  Department of Physics and CENIDE, University of Duisburg-Essen, Duisburg, Germany}

\date{Date: November 12, 2015}

\abstract{The current flow along the boundary of graphene stripes in a perpendicular magnetic field is studied
  theoretically by the nonequilibrium Green's function method. In the case of specular reflections at the
  boundary, the Hall resistance shows equidistant peaks, which are due to classical cyclotron motion. When the
  strength of the magnetic field is increased, anomalous resistance oscillations are observed, similar to
  those found in a nonrelativistic 2D electron gas [New. J. Phys. 15:113047 (2013)]. Using a simplified model,
  which allows to solve the Dirac equation analytically, the oscillations are explained by the interference
  between the occupied edge states causing beatings in the Hall resistance. A rule of thumb is given for the
  experimental observability. Furthermore, the local current flow in graphene is affected significantly by the
  boundary geometry. A finite edge current flows on armchair edges, while the current on zigzag edges vanishes
  completely. The quantum Hall staircase can be observed in the case of diffusive boundary scattering. The
  number of spatially separated edge channels in the local current equals the number of occupied Landau
  levels. The edge channels in the local density of states are smeared out but can be made visible if only a
  subset of the carbon atoms is taken into account.}

\maketitle

\section{Introduction} \label{sec:intro}

Nowadays, graphene is maybe the most studied material in condensed matter physics because of its numerous
exceptional properties and their potential technological applications, see \cite{Geim2007, Geim2009,
  CastroNeto2009, Avouris2010, Novoselov2012, Katsnelson2012} and references therein for an overview. In
particular the electronic transport of charge carriers in graphene is of enormous interest due to the promise
of novel electronic devices like foldable displays and high-frequency transistors \cite{Novoselov2012}. It has
also been shown recently that strain and deformation of a graphene stripe, as caused by the absorption of
atoms for example \cite{Jalbout2013}, give rise to a strong pseudo-magnetic field \cite{Guinea2010, Levy2010},
which affects significantly the current flow \cite{Stegmann2015}.

\begin{figure}[t]
  \centering
  \vspace*{-2mm}
  \includegraphics[scale=0.43]{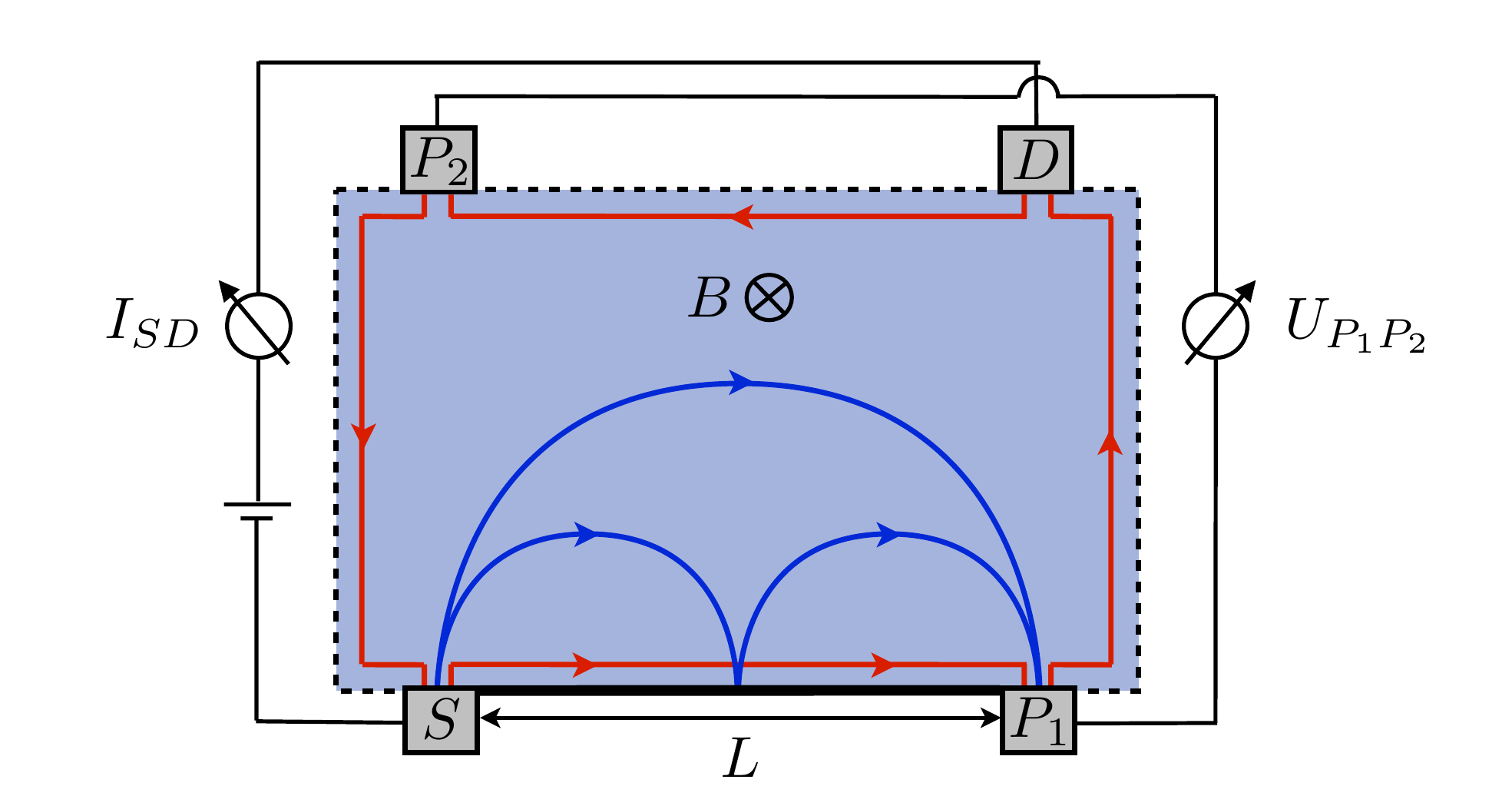}
  \caption{Magnetotransport in graphene stripes is studied. Electrons are injected at the source $S$ and
    focussed by a perpendicular magnetic field $B$ onto the contact $P_1$. We calculate the current $I_{SD}$
    between source $S$ and drain $D$ as well as the voltage drop $U_{P_1P_2}$ between the voltage probes $P_1$
    and $P_2$. Using these quantities, we study the generalized Hall resistance $R_{xy}= U_{P_1P_2}/I_{SD}$ as
    a function of $B$. Cyclotron orbits at low magnetic field are sketched by the blue trajectories. The edge
    channel transport of the quantum Hall effect at high magnetic field is indicated by red lines.}
  \label{fig:1}
\end{figure}

In this paper, we study magnetotransport along the edges of graphene stripes. As sketched in \fig{1},
electrons are injected coherently at one point $S$ on the boundary of the graphene stripe and focussed by a
homogeneous perpendicular magnetic field $B$ onto another point $P_1$ on that boundary. In the classical
regime (blue trajectories) resonances are observed, if a multiple of the cyclotron diameter equals the
distance between the injecting and collecting point contacts \cite{Tsoi1974, Tsoi1999}. For large Fermi
wavelength and long phase coherence length, additional interference effects appear. This regime of
\textit{coherent electron focusing} has been studied for the first time by van Houten et al. in a
nonrelativistic two-dimensional electron gas (2DEG) \cite{Houten1989} but it has become a topic of current
interest again, since the first focusing experiments in graphene have become possible \cite{Taychatanapat2013,
  Calado2014}. The magnetic focusing in graphene pn junctions has been studied theoretically
\cite{Milovanovic2014b} and snake states at such a pn interface have been predicted \cite{Milovanovic2013,
  Milovanovic2014}. It has also been suggested to study by coherent electron focusing the structure of
graphene edges \cite{Rakyta2010}. Recently, also the effects of disorder \cite{Maryenko2012} and spin-orbit
interaction \cite{Usaj2004, Rokhinson2004, Dedigama2006, Reynoso2008, Kormanyos2010} have been investigated in
a nonrelativistic 2DEG. On the other hand, graphene stripes in a strong magnetic field show the quantum Hall
effect \cite{Novoselov2005, Zhang2005}, which is explained by the transport through \textit{edge channels}
along the boundary of the system, see the red lines in \fig{1}.

Here, we study theoretically the system properties from the classical to the quantum regime. In particular, we
discuss the novel effects which emerge, when the two regimes are bridged by suitable system parameters. If the
scattering at the boundaries is specular, we observe in this intermediate regime anomalous resistance
oscillations, which are neither periodic in $ B $ (classical cyclotron motion) nor periodic in $1/B$ (quantum
Hall effect). Using a simplified model, which allows to solve the Dirac equation analytically, we explain
these oscillations by the interference of the occupied edge channels. These anomalous resistance oscillations
have been reported recently in a nonrelativistic 2DEG \cite{Stegmann2013}. Beyond this, in graphene the local
current flow is affected significantly by the boundary geometry. We show that on armchair edges a finite edge
current is present, while on zigzag edges the current is shifted to the interior of the stripe and vanishes
exactly on the edge. We also give a rule of thumb for the experimental observability of these effects. The
quantum Hall staircase is observed in the case of diffusive scattering at the boundaries, which guarantees
that the phase coherence length is shorter than the relevant geometric lengths (point contact distance, system
size). We show that the number of spatially separated edge channels in the local current equals the number of
occupied Landau levels. In the local density of states (LDOS) the edge channels are smeared out but can be
made visible if only a subset of the carbon atoms is considered.

\vspace*{-2mm}
\section{System} \label{sec:system}

We study a graphene stripe with a size of $140 \un{nm} \times 90 \un{nm}$, see \fig{1}. Metallic
contacts with a width of $3 \un{nm}$ are attached at the edges of the stripe separated by a distance
of $L= 110 \un{nm}$ (measured between the middle of the contacs). We calculate the current $I_{SD}$
flowing between source $S$ and drain $D$ due to an infinitesimal bias voltage, as well as the
voltage drop $U_{P_1P_2}$ between the voltage probes $P_1$ and $P_2$. Using these quantities, we
study the generalized Hall resistance $R_{xy}= U_{P_1P_2}/I_{SD}$ as a function of an homogeneous
perpendicular magnetic field $\vec{B}= -B \vec{e}_z$.  The Fermi energy is set to
$\mu= 0.06t = 168 \un{meV}$, where $t=2.8 \un{eV}$ is the coupling between nearest neighboring
carbon atoms at a distance $a=0.142 \un{nm}$ \cite{CastroNeto2009}. In the experiment, usually the
electron density is constant while the chemical potential is oscillating. However, this would only
slightly displace the transitions between the Hall plateaus and would not qualitatively change our
results, see also \cite{Gagel1998}. For simplicity, we do not take into account the Zeeman spin
splitting, see remarks in section~\ref{sec:exp-obs}. We also assume that the influence of the
temperature is negligible and thus, set it to zero. We consider graphene stripes with a zigzag
boundary in between $S$ and $P_1$, as well as with an armchair boundary, see the inset of
\fig{4}. Other possible edge reconstructions, see e.g. \cite{Dubois2010, Hawkins2012,
  Ihnatsenka2013}, are not considered here. Note that the orientation of the graphene lattice with
respect to the used coordinate system is not changed in the two stripes. Armchair edges run along
the $x$-axis, whereas zigzag edges are oriented along the $y$-axis, see the inset of \fig{4}.

\vspace*{-2mm}
\section{Calculations} \label{sec:calculations}

In this section, we begin with a short introduction into the nonequilibrium Green's function (NEGF) method,
which is applied to study quantitatively the magnetotransport in graphene stripes. Afterwards, we solve the
Dirac equation in a magnetic field for a graphene sheet bounded by a single infinite potential wall. This
simplified model will help us to get insight into the results of the Green's function calculations.

\vspace*{-1mm}
\subsection{The nonequilibrium Green's function method} \label{sec:negf}

We start from the tight-binding Hamiltonian of the gra\-phene stripe
\begin{equation}
  \label{eq:1}
  H= -t \sum_{\braket{ij}} \ket{\phi_j^A} \bra{\phi_i^B} + \tx{H.c.},  
\end{equation}
where $\braket{ij}$ means nearest neighbors at a distance $a $ with coupling $t$.  The
$\ket{\phi^{A/B}_i}$ are the $p_z$ orbitals of the carbon atoms on sublattice A and B, respectively,
see the red and black marked atoms in the inset of \fig{4}. In graphene nanoribbons it can be
necessary to take into account also the interaction to second and third nearest neighbors
\cite{Cresti2008d, Boykin2011, Chang2012}. However, for the graphene stripes studied here, it is
sufficient to consider only nearest neighbors, as our main results remain qualitatively unchanged if
also second and third nearest neighbors are taken into account.

In the tight-binding Hamiltonian the effect of the magnetic field $\vec{B}$ is
taken into account by the Peierls substitution \cite{Feynman1963}
\begin{equation}
  \label{eq:2}
  t_{ij}(B)= t_{ij}(B=0) \E^{\I \frac{e}{h} \int \d \vec{l} \cdot \vec{A}},
\end{equation}
where $\vec{A}$ is the vector potential of the magnetic field. The path integral is along the straight
connection between the position of carbon atom $i$ and $j$.

The Green's function of the graphene stripe is defined as \cite{Datta1997, Datta2005, Datta2012}
\begin{equation}
  \label{eq:3}
  G= \Big[E -H -\textstyle{\sum_{k=1}^{N_c} \Gm_k} \Big]^{-1},
\end{equation}
where $E$ is the energy of the charge carriers. The influence of each of the $N_c$ contacts is taken into
account by an imaginary self-energy
\begin{equation}
  \label{eq:4}
  \Gm_k= -\I \eta \sum_{\vec{r}_i} \ket{\vec{r}_i}\bra{\vec{r}_i}
\end{equation}
with broadening $\eta= 1.25t= 3.5\un{eV}$, representing metallic contact regions. The sum is over all carbon
atoms which are coupled to the same contact $k$.

The transmission from contact $j$ to contact $i$ is then given by
\begin{equation}
  \label{eq:5}
  T_{ij}= 4\Tr{\Im{\Gm_i}G\Im{\Gm_j}G^\dagger}.
\end{equation}
and the total current at the $i$th contact reads
\begin{equation}
  \label{eq:6}
  I_i= \frac{2e}{h} \sum_j T_{ij} \lr{\mu_j-\mu_i},
\end{equation}
where $\mu_{i/j}$ is the chemical potential of contact $i$ and $j$, respectively. The generalized Hall
resistance is then given by
\begin{equation}
  \label{eq:7}
  R_{xy}= \frac{U_{P_1P_2}}{I_{SD}}
  = \frac{h}{2e^2}\,\frac{\sum_j \lr{\mathcal{R}_{P_1j}-\mathcal{R}_{P_2j}}T_{jS}}{T_{DS} +\sum_{ij}T_{Di}\mathcal{R}_{ij}T_{jS}},
\end{equation}
where
\begin{equation}
  \label{eq:8}
  \mathcal{R}^{\,-1}_{ij}=
  \begin{cases}
    -T_{ij}  & i \neq j,\\
    \textstyle{\sum_{k \neq i} T_{ik}}  & i = j.
  \end{cases}
\end{equation}
The sums in \eq{7} are over the contacts with unknown chemical potential, whereas the sum in \eq{8} is over
all contacts including source and drain. 

The local current of electrons, which originate from the source with energy $\mu$ and which flow from atom $j$
to the neighboring atom $i$, is given by \cite{Caroli1971, Cresti2003}
\begin{equation}
  \label{eq:9}
  I_{ij}= \frac{2e}{\hbar} \Im{H_{ji}^*A_{ji}^S},
\end{equation}
where the $H_{ij}$ are the matrix elements of the Hamiltonian~\eq{1}. The spectral function for electrons from
the source is defined as
\begin{equation}
  \label{eq:10}
  A^S= -\frac{2}{\pi}G\Im{\Gm_S}G^+.
\end{equation}
The diagonal elements of the spectral function give the local density of states (LDOS), which is accessible to
these electrons.

Finite system size effects, such as standing waves between the system boundaries, would distort the
magnetotransport strongly. Therefore, diffusive boundaries are used at those edges, which are not
important for the focusing experiment, see the dashed edges in \fig{1}. Diffusive boundaries are
implemented mathematically by additional virtual voltage probes, which randomize phase and momentum
of the charge carriers and thus, suppress standing waves in the system. The chemical potential of
the virtual reservoirs is determined by the condition that no charge carriers can be gained or lost
at a virtual reservoir (current conservation constraint), see \cite{Zilly2009, Stegmann2012,
  Stegmann2013} for details.

\subsection{Dirac equation in a magnetic field} \label{sec:dirac}

For Fermi energies close to $E=0$, the physics of graphene takes place at two points $\vec{K}$ and $\vec{K}'$
in momentum space. At these points the dispersion relation is linear $E(\vec{k})= \hbar v_F \Abs{\vec{k}}$ and
the charge carriers behave as relativistic massless particles described by the Dirac Hamiltonian
\begin{equation}
  \label{eq:11}
  H_{\vec{K}/\vec{K}'}= v_F
  \begin{pmatrix}
    0 & p_x \mp \I p_y\\
    p_x \pm \I p_y & 0
  \end{pmatrix}
\end{equation}
with $v_F= 3 a t/ 2 \hbar$. Note that in order to derive from the tight-binding Hamiltonian~\eq{1} the Dirac
Hamiltonian~\eq{11} in its common notation \cite{Geim2007, CastroNeto2009, Katsnelson2012}, we also applied an
unitary transformation
\begin{equation}
  \label{eq:12}
  U= \begin{pmatrix} 1&0\\ 0&\I \end{pmatrix}.
\end{equation}

The effect of the magnetic field is taken into account by minimal gauge invariant coupling
$\vec{p} \to \vec{p} -e \vec{A}$, where $\vec{A}= B y \vec{e}_x$ is the vector potential chosen for the
armchair stripe, see \fig{4} (bottom). To solve the Dirac equation, we insert the two linear equations into
each other, keeping in mind that $\left[ p_x, p_y \right]= -\I e B \hbar$. By means of the ansatz
$\psi_{B}(\vec{r})= \E^{\I k x} \chi_B(y)$, we obtain the Schr\"odinger equation of a harmonic oscillator
\begin{equation}
  \label{eq:13}
  \widetilde{E} \, \chi_B(y) = \biggl[ \frac{p_y^2}{2m} +\frac{1}{2}m\om_c^2\lr{y-y_k}^2 \biggr]\chi_B(y),
\end{equation}
which is shifted by $y_k= \ell_B^2 k$ and rescaled in energy
$\widetilde{E} = \frac{E^2}{2m v_F^2} + \frac{\hbar \om_c}{2}$ with $\om_c= \frac{eB}{m}$ and
$\ell_B^2= \frac{\hbar}{eB}$. Thus, the eigenenergies are given by
\begin{equation}
  \label{eq:14}
  E_{\nu}= \pm \sqrt{2 m v_F^2 \, \hbar \om_c \, \nu}, \qquad \nu \geq 0.
\end{equation}
The eigenstates on sublattice~B read $\chi_B \sim \mc{D}_{\nu}\bigl(\frac{y-y_k}{\ell_B} \bigr)$, where
$ \mc{D}_{\nu}(y) \equiv D_{\nu}(\sqrt{2} y)/\sqrt{\nu!} $ are rescaled parabolic cylinder functions
\cite{Abramowitz1965}. The eigenstates on sublattice~A follow directly from $\chi_B$, the Dirac equation and
the recursion relation $\lr{\pd_y + y}\mc{D}_{\nu}(y)= \sqrt{2\nu} \mc{D}_{\nu-1}(y)$, see
\cite{Abramowitz1965}. The solution of the Dirac equation at the $\vec{K}'$ valley can be obtained easily by
interchanging the two sublattices, see \eq{11}. The eigenenergy spectrum is unchanged and hence, twofold
degenerate. Thus, the eigenfunctions are given by
\begin{subequations}
  \label{eq:15}
  \begin{align}
    \psi_{\vec{K}}(\vec{r})&= c_{\nu} \, \E^{\I k x}
                             \begin{pmatrix}
                               \mp \mc{D}_{\nu-1}(\xi)\\[1ex]
                               \I \mc{D}_{\nu}(\xi)
                             \end{pmatrix},\\[1ex]
    \psi_{\vec{K}'}(\vec{r})&= c_{\nu} \, \E^{\I k x}
                              \begin{pmatrix}
                                \mc{D}_{\nu}(\xi)\\[1ex]
                                \mp \I \mc{D}_{\nu-1}(\xi)
                              \end{pmatrix},
  \end{align}
\end{subequations}
where $c_{\nu}$ is a normalization constant, $\xi \equiv \lr{y-\ell_B^2 k}/\ell_B$ and
$ \mc{D}_{x<0} \equiv 0 $. The different signs of the eigenstates correspond to the signs of the
eigenenergies. We also applied the unitary transformation \eq{12} in order to get the correct phase between
the wavefunctions on the sublattices. For the zigzag stripe, see \fig{4} (top), we choose the vector potential
$\vec{A}= -B x \vec{e}_y$, to get the eigenstates
\begin{subequations}
  \label{eq:16}
  \begin{align}
    \psi_{\vec{K}}(\vec{r})&= c_{\nu} \, \E^{\I k y}
                             \begin{pmatrix}
                               \mp \I \mc{D}_{\nu-1}(\zeta)\\[1ex]
                               \I \mc{D}_{\nu}(\zeta)
                             \end{pmatrix},\\[1ex]
    \psi_{\vec{K}'}(\vec{r})&= c_{\nu} \, \E^{\I k y}
                              \begin{pmatrix}
                                \mc{D}_{\nu}(\zeta)\\[1ex]
                                \pm \mc{D}_{\nu-1}(\zeta),
                              \end{pmatrix},
  \end{align}
\end{subequations}
where $\zeta \equiv \lr{x+\ell_B^2 k}/\ell_B$. In an infinitely extended system, the index $\nu$ has to be an
integer $n=0,1,2\hdots$ (Landau level index), because the eigenfunctions have to be normalizable. In this
case, the parabolic cylinder functions can be simplified by $\mc{D}_n(y)= \E^{-y^2/2} H_n(y)/\sqrt{2^n n!}$,
where $H_n(y)$ are the Hermite polynomials.

\subsection{Dirac equation with an edge in a magnetic field} \label{sec:dirac-equat-bound}

\begin{figure}[htb]
  \centering
  \vspace*{-3mm}
  \includegraphics[scale=0.82]{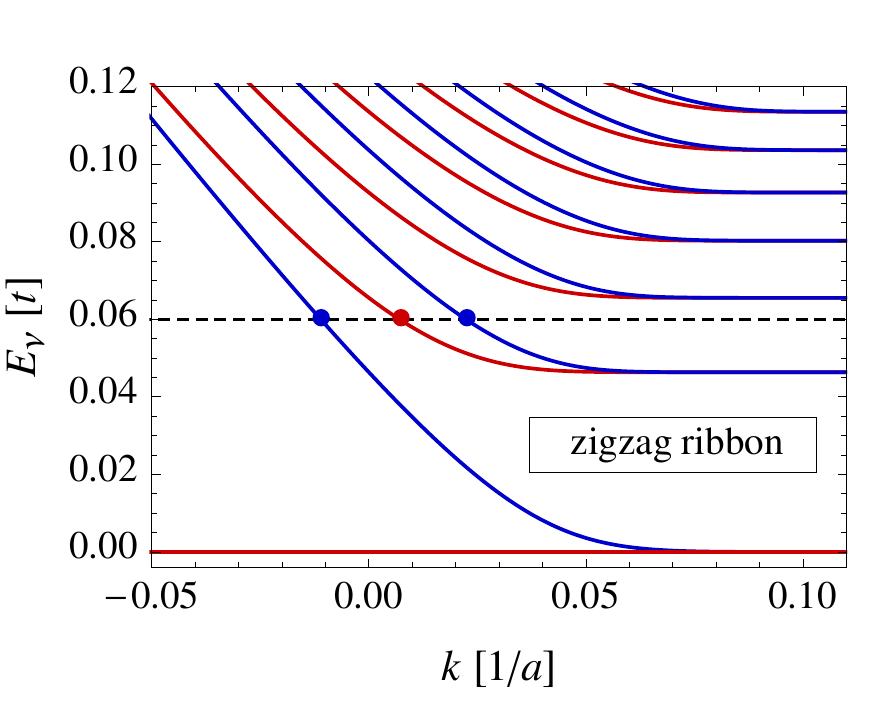}\\[-2mm]
  \includegraphics[scale=0.82]{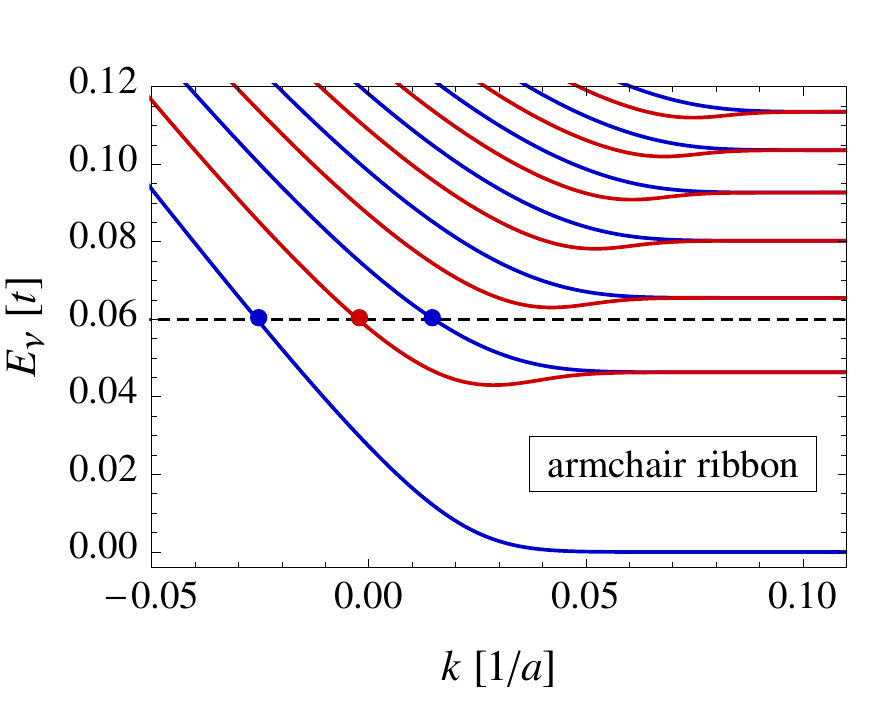}
  \caption{Energy bands of graphene bounded by an edge in a magnetic field of $B= 15.6\un{T}$.  At large $k$
    we observe discrete Landau levels. When $k$ is decreased, the energy bands are bent upwards and their
    degeneracy is lifted. In the case of a zigzag edge (top) the red curve indicates solutions at the
    $\vec{K}$ valley, while the blue curve gives solutions at the $\vec{K}'$ valley. In the case of an
    armchair edge (bottom) the valleys are mixed, which leads to two sets of solutions indicated by the blue
    and red curve. The occupied edge states at the Fermi energy (dashed horizontal line) are marked by blue
    and red dots.}
  \label{fig:2}
\end{figure}

To understand the magnetotransport in graphene stripes, we solve the Dirac equation bounded by an edge
under the effect of a magnetic field. In general, the solution of the Dirac equation is given by a linear
combination of the solutions at both valleys
\begin{equation}
  \label{eq:17}
  \Psi(\vec{r})= c_1 \, \E^{\I \vec{K} \cdot \vec{r}} \psi_{\vec{K}}(\vec{r}) +c_2 \, \E^{\I \vec{K}' \cdot \vec{r}} \psi_{\vec{K}'}(\vec{r}),
\end{equation}
where $c_1$ and $c_2$ are complex constants. At the zigzag edge, we obtain by means of \eq{16}
\begin{align}
  \label{eq:18}
  \Psi_{\tx{zz}}(\vec{r})= \E^{\I\lr{\frac{2\pi}{3}x+ky}}\biggl[ 
    &c_1 \, \E^{\I \frac{2\pi}{3\sqrt{3}} y}
    \begin{pmatrix}
      \mp \I \mc{D}_{\nu-1}(\zeta) \\
      \I \mc{D}_{\nu}(\zeta)
    \end{pmatrix} \notag \\
    +&c_2 \, \E^{-\I \frac{2\pi}{3\sqrt{3}} y}
    \begin{pmatrix}
       \mc{D}_{\nu}(\zeta)\\
       \pm \mc{D}_{\nu-1}(\zeta)
    \end{pmatrix}
\biggr].
\end{align}
As only carbon atoms of one sublattice appear at a zigzag edge, see the inset of \fig{4} (top), the wave
function has to vanish only on one of the two sublattices. The condition $\Psi_A(x=0)=0 $ leads to the two
solutions
\begin{subequations}
  \label{eq:19}
  \begin{align}
    &c_1=1, \: c_2=0: \quad \mc{D}_{\nu-1}(\ell_B k) \ovs{!}{=} 0,\\
    &c_1=0, \: c_2=1: \quad \mc{D}_{\nu}(\ell_B k) \ovs{!}{=} 0.
  \end{align}
\end{subequations}
Thus, for given $\ell_B k=\sqrt{\hbar k^2 /e B}$ the index $\nu$ is determined by the zeros of the rescaled
parabolic cylinder functions. The first set of solutions is located at the $\vec{K}$ valley, whereas the
second set is located at the $\vec{K}'$ valley. The resulting energy bands \eq{14} are depicted in \fig{2}
(top). At large $k$ the discrete Landau levels for integer values of $\nu= n$ can be observed. The distance of
these Landau levels decreases with $\sqrt{n}$. When the apex of the parabola $y_k=\ell_B^2 k$ approaches the
wall by decreasing $k$, the energy bands are bent upwards and their degeneracy is lifted. Also a
dispersionless state $E_{\nu=0}=0$ at the $\vec{K}$ valley can be seen. The occupied edge states at the Fermi
energy (dashed horizontal line) are indicated by dots.

At an armchair edge, we obtain by means of \eq{15}
\begin{align}
  \label{eq:20}
  \Psi_{\tx{ac}}(\vec{r})= \E^{\I\lr{\frac{2\pi}{3} +k}x}\biggl[ 
    &c_1 \, \E^{\I \frac{2\pi}{3\sqrt{3}} y}
    \begin{pmatrix}
      \mp \mc{D}_{\nu-1}(\xi) \\
      \I \mc{D}_{\nu}(\xi)
    \end{pmatrix} \notag \\
    +&c_2 \, \E^{-\I \frac{2\pi}{3\sqrt{3}} y}
    \begin{pmatrix}
      \mc{D}_{\nu}(\xi)\\
      \mp \I \mc{D}_{\nu-1}(\xi)
    \end{pmatrix}
  \biggr].
\end{align}
As at armchair edges both sublattices appear, see the inset of \fig{4} (bottom), the wave function has to
vanish on both of them. The condition $\Psi_A(y=0)= \Psi_B(y=0)=0$ requieres that the coefficient determinant
of the linear equation system for $c_1$ and $c_2$ vanishes
\begin{equation}
  \label{eq:21}
  \mc{D}_{\nu-1}(-\ell_B k)\mp \mc{D}_{\nu}(-\ell_B k)=0
\end{equation}
and leads to the solutions
\begin{equation}
  \label{eq:22}
  c_1=1, \quad  c_2=\pm 1.
\end{equation}
Thus, at an armchair edge both valleys are intermixed. The two eigenenergy bands in \fig{2} (bottom) show not
only that their degeneracy is lifted in vicinity of the edge but also shallow valleys, which are not present
at a zigzag edge. The solution of the Dirac equation at zigzag and armchair edges in a magnetic field can also
be found in \cite{Brey2006_2, Abanin2007, Delplace2010, Wang2011}.

\section{Results and discussion} \label{sec:results}

\subsection{Density of states} \label{sec:dens-stat}

\begin{figure}[htb]
  \centering
  \vspace*{-3mm}
  \includegraphics[scale=0.82]{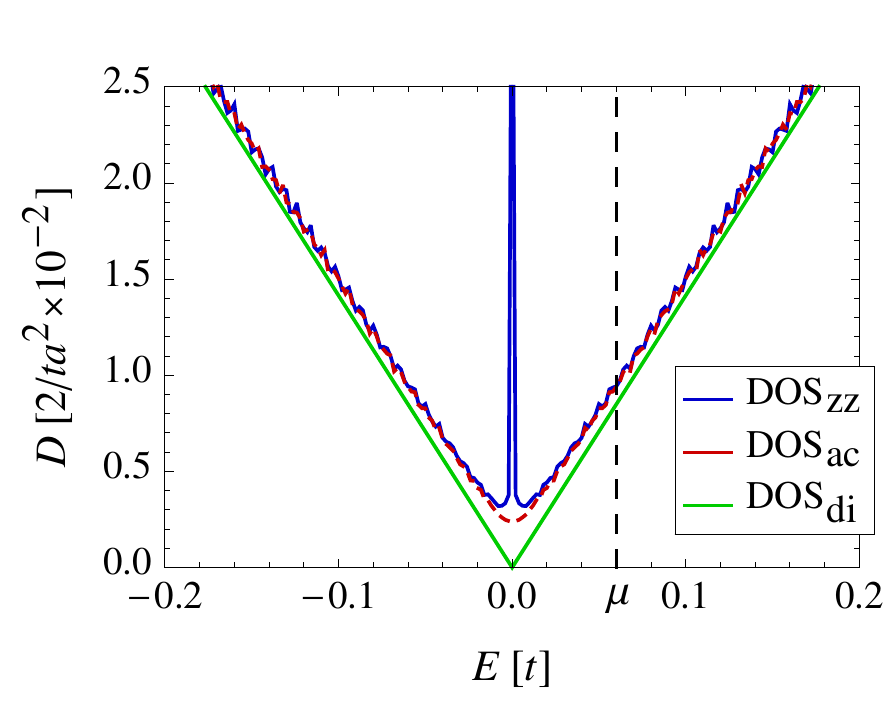}
  \caption{Average DOS in the studied graphene stripes. The blue curve gives the DOS in the zigzag stripe,
    while the red curve gives the DOS in the armchair stripe, see the inset of \fig{4}. The DOS of both
    stripes agrees well with the DOS from the Dirac Hamiltonian (green curve). The peak in the DOS of the
    zigzag stripe at $E=0$ is caused by the dispersionless surface state, which does not exist in armchair
    stripes. However, also the armchair stripe has a nonzero DOS at $E=0$, because of contact induced
    states. The dashed vertical line indicates the Fermi energy.}
  \label{fig:3}
\end{figure}

Let us discuss briefly the density of states (DOS) in the studied devices, which is shown in \fig{3}. For
energies $E>0.02t$ the DOS in the zigzag and the armchair stripe agree well with the DOS
$D_{\tx{di}}(E)= \frac{16 \Abs{E}}{9 \pi ta^2}$ from the Dirac Hamiltonian. However, the zigzag stripe shows a
distinct peak at $E=0$, which cannot be observed in the case of an armchair stripe. This peak can be
attributed to the dispersionless state shown in \fig{2} (top), which does not contribute to electron transport
and is located on the surface of the stripe. A surface state is possible at a zigzag edge, because only carbon
atoms of a single sublattice appear there. Thus, at the edge the wave function has to vanish only on one
sublattice, while the surface state resides on the other sublattice. At an armchair edge atoms from both
sublattices appear and a surface state is not possible, see \cite{Nakada1996, Wakabayashi2009, CastroNeto2009,
  Hawkins2012} for details. However, in \fig{3} the DOS of the armchair stripe is also nonzero at $E=0$. These
are states induced by the contacts \cite{Golizadeh-Mojarad2009}, which contribute to the observed finite
conductivity of graphene at the Dirac points \cite{Novoselov2004, Novoselov2005, Zhang2005, Tan2007,
  Chen2008}. As a consequence of this, at $\mu=0.06t=168 \un{meV}$, the carrier densities in the zigzag stripe
$n_{\tx{zz}}= 3.7 \cdot 10^{12} \un{cm}^{-2}$ and the armchair stripe
$n_{\tx{ac}}= 3.3 \cdot 10^{12} \un{cm}^{-2}$ are somewhat higher than expected from the linear DOS of the
Dirac Hamiltonian $n_{\tx{di}}= 2.5 \cdot 10^{12} \un{cm}^{-2}$.

\begin{figure}[htb]
  \centering
  \vspace*{-3mm}
  \includegraphics[scale=0.82]{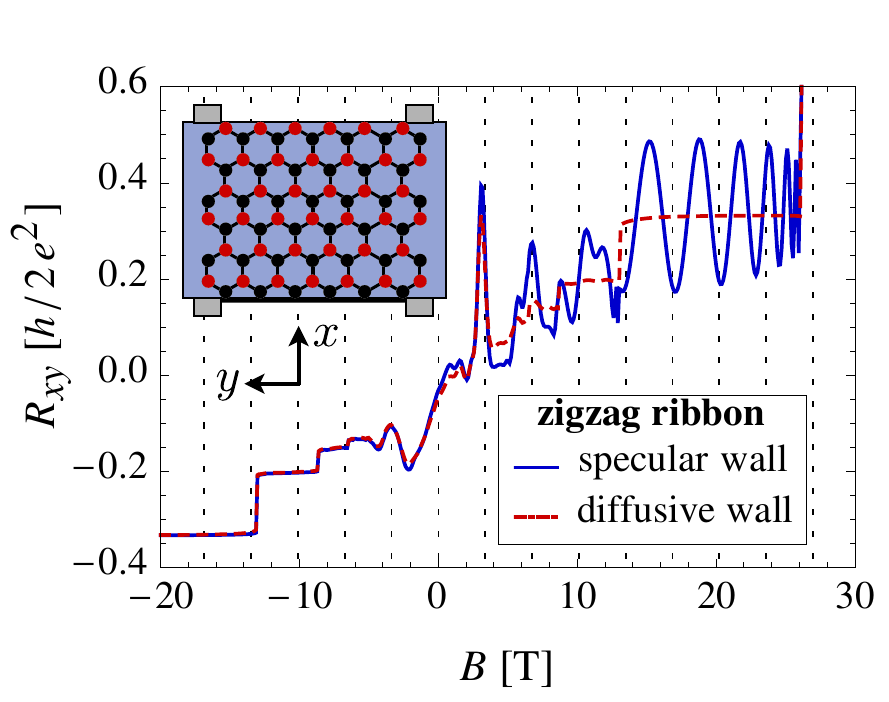}\\[-2mm]
  \includegraphics[scale=0.82]{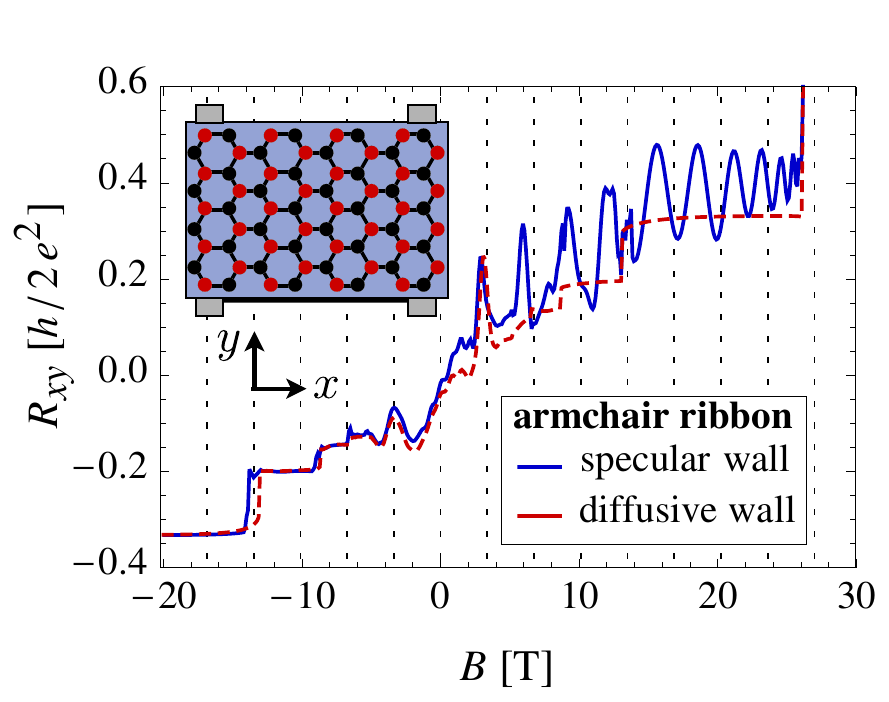}
  \caption{Hall resistance $R_{xy}$ as a function of the magnetic field $B$ for the zigzag stripe (top) and
    the armchair stripe (bottom). The blue curve gives $R_{xy}$ in the case of specular reflections at the
    boundary between $S$ and $P_1$, whereas for the red curve the scattering at this boundary is
    diffusive. The Hall resistance starts with peaks, which can be understood by classical cyclotron orbits
    \eq{23}, see the dashed vertical lines. In a strong magnetic field $B> 10\un{T}$, we observe superimposed
    upon the quantum Hall plateaus anomalous resistance oscillation, which cannot be explained by cyclotron
    orbits.}
  \label{fig:4}
\end{figure}

\begin{figure*}[htb] 
  \centering
  \vspace*{-2mm}
  \includegraphics[scale=0.5]{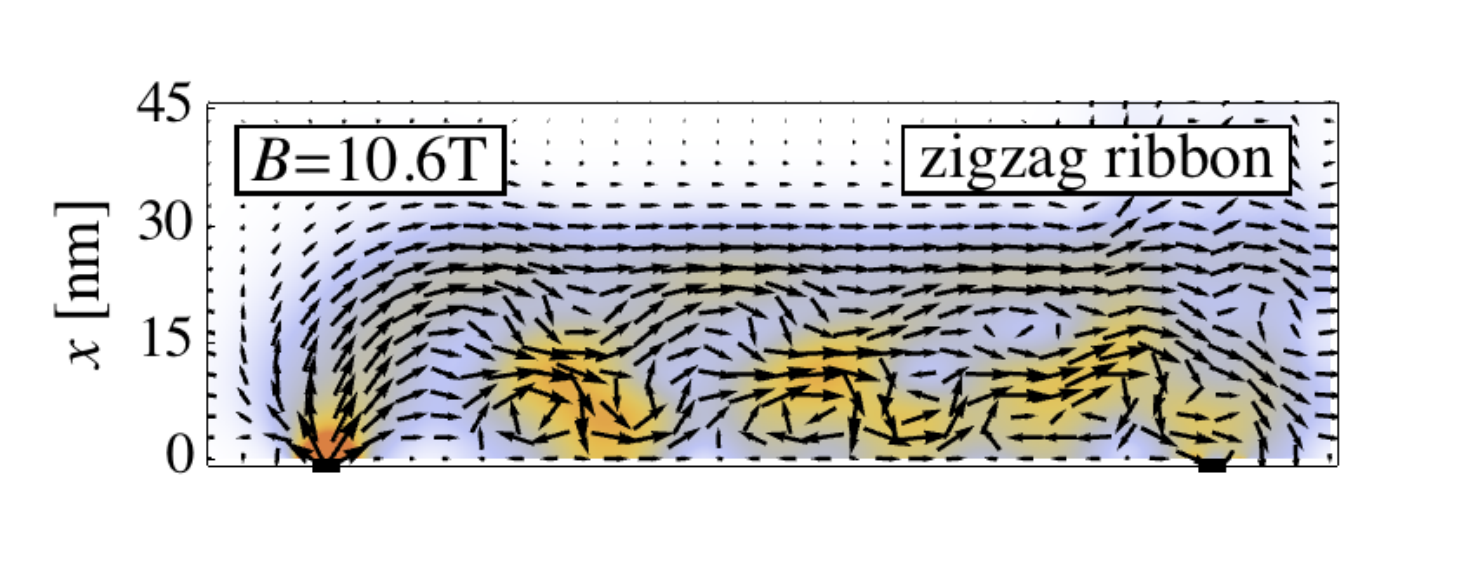}  \includegraphics[scale=0.5]{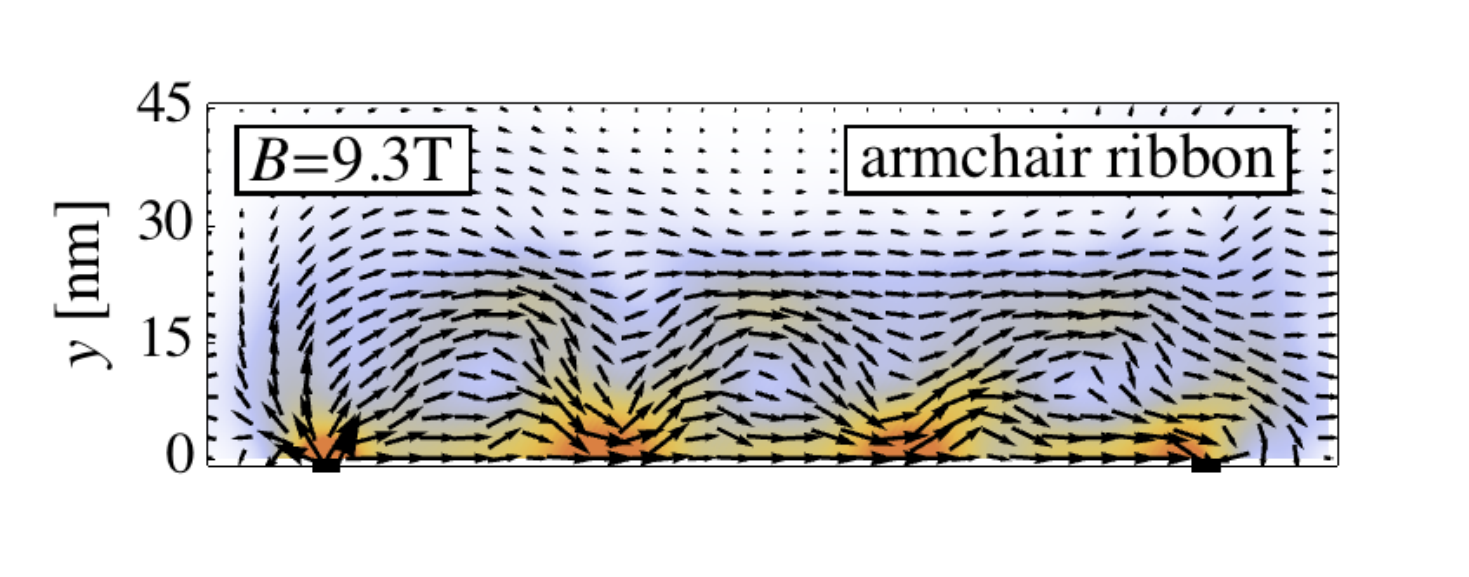}\\[-4mm]
  \includegraphics[scale=0.5]{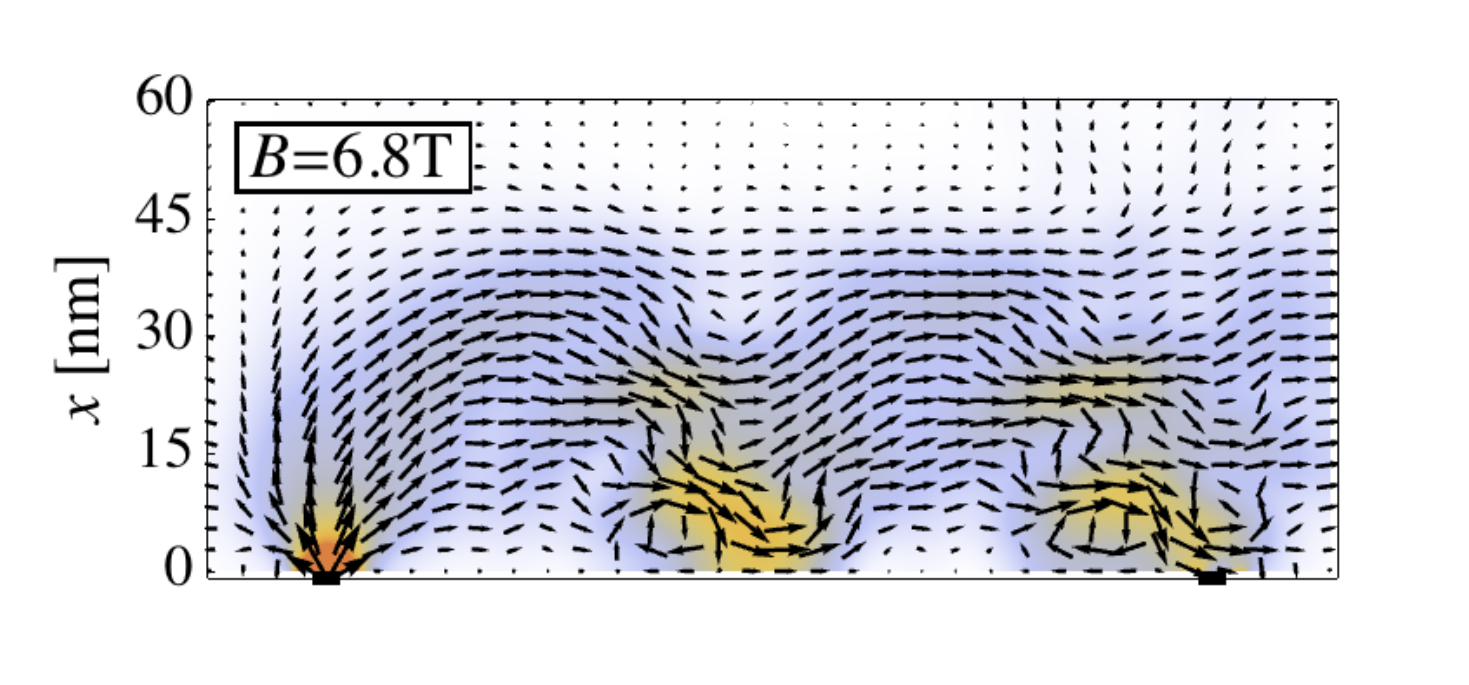}  \includegraphics[scale=0.5]{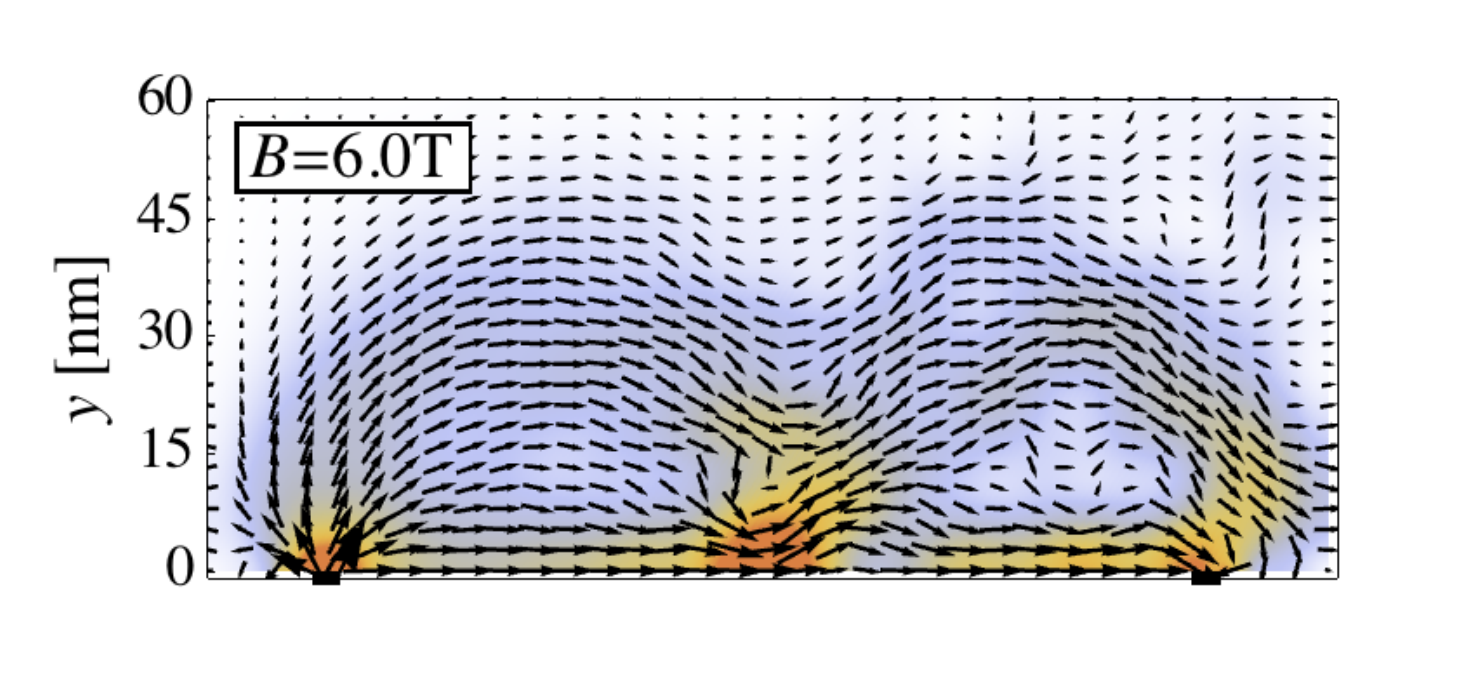}\\[-4mm]
  \includegraphics[scale=0.5]{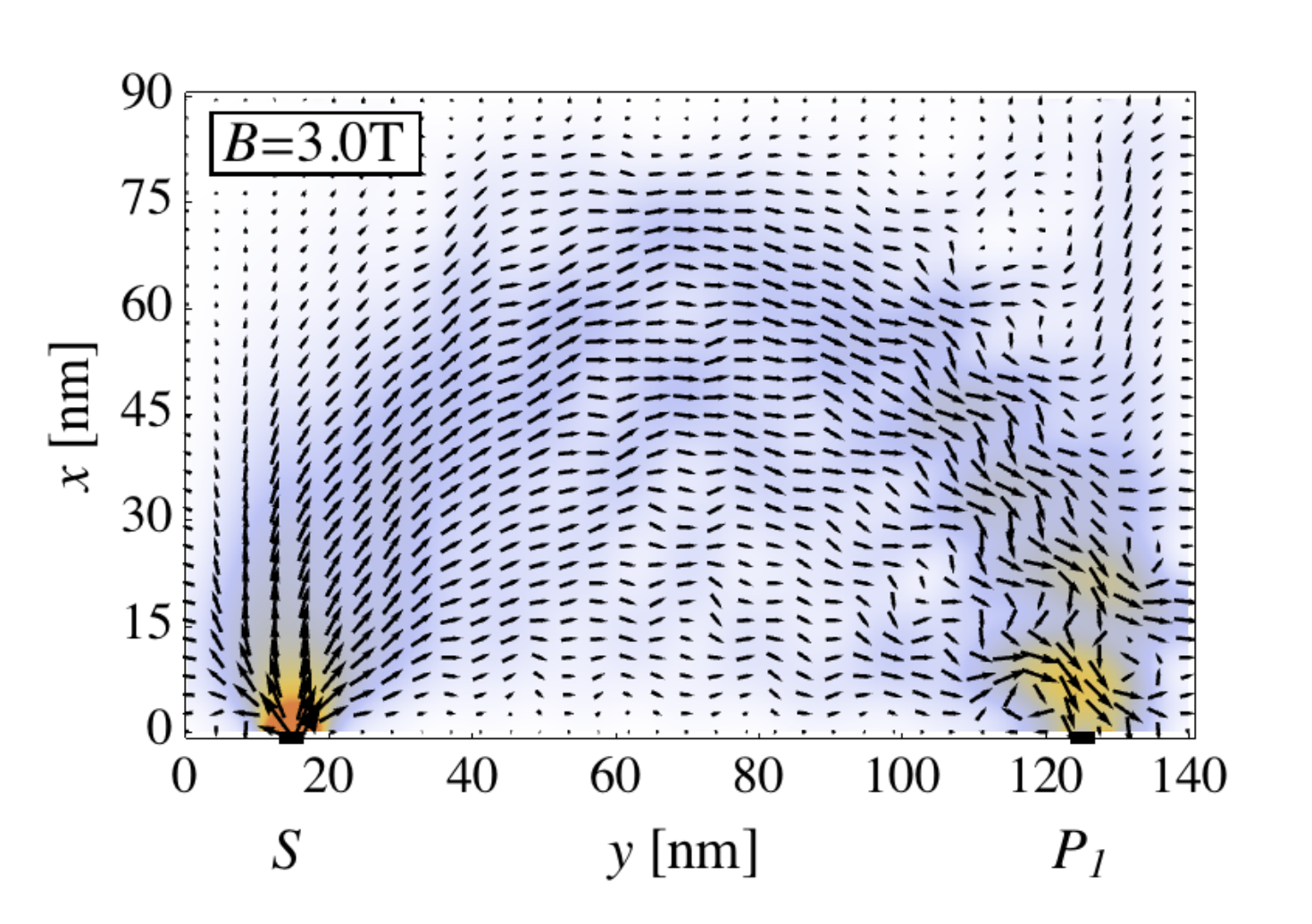}  \includegraphics[scale=0.5]{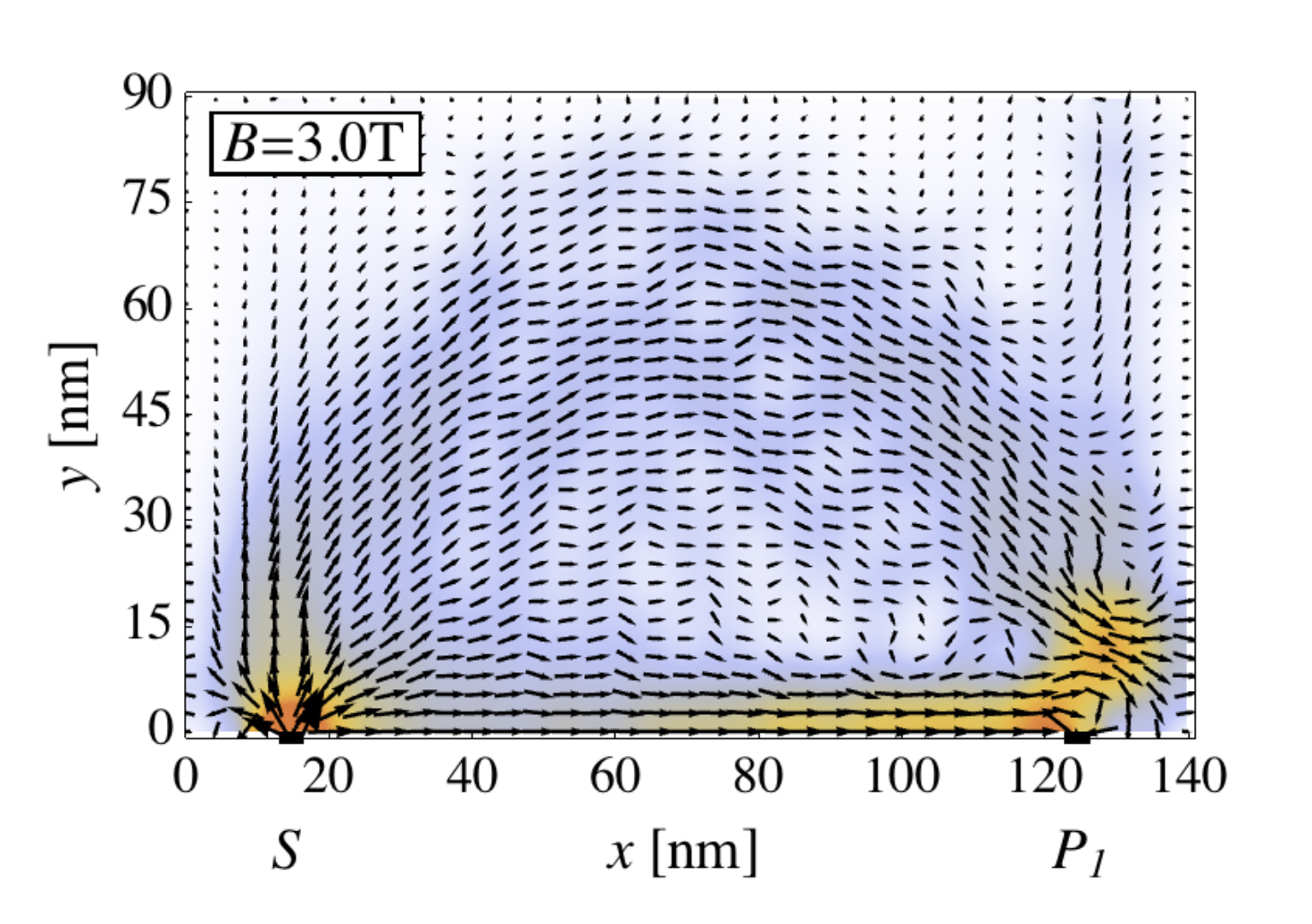}
  \caption{Local current (arrows) and LDOS (shading) of electrons originating from $S$ with energy $\mu$. In
    the zigzag stripe (left column) and the armchair stripe (right column) cyclotron orbits can be clearly
    seen. At the armchair edge a distinct edge current can be observed, which is not present at the zigzag
    edge. Note that the shown local current and the LDOS have been averaged over the honeycomb cells.}
  \label{fig:5}
\end{figure*}

\begin{figure*}[htb] 
  \centering
  \includegraphics[scale=0.5]{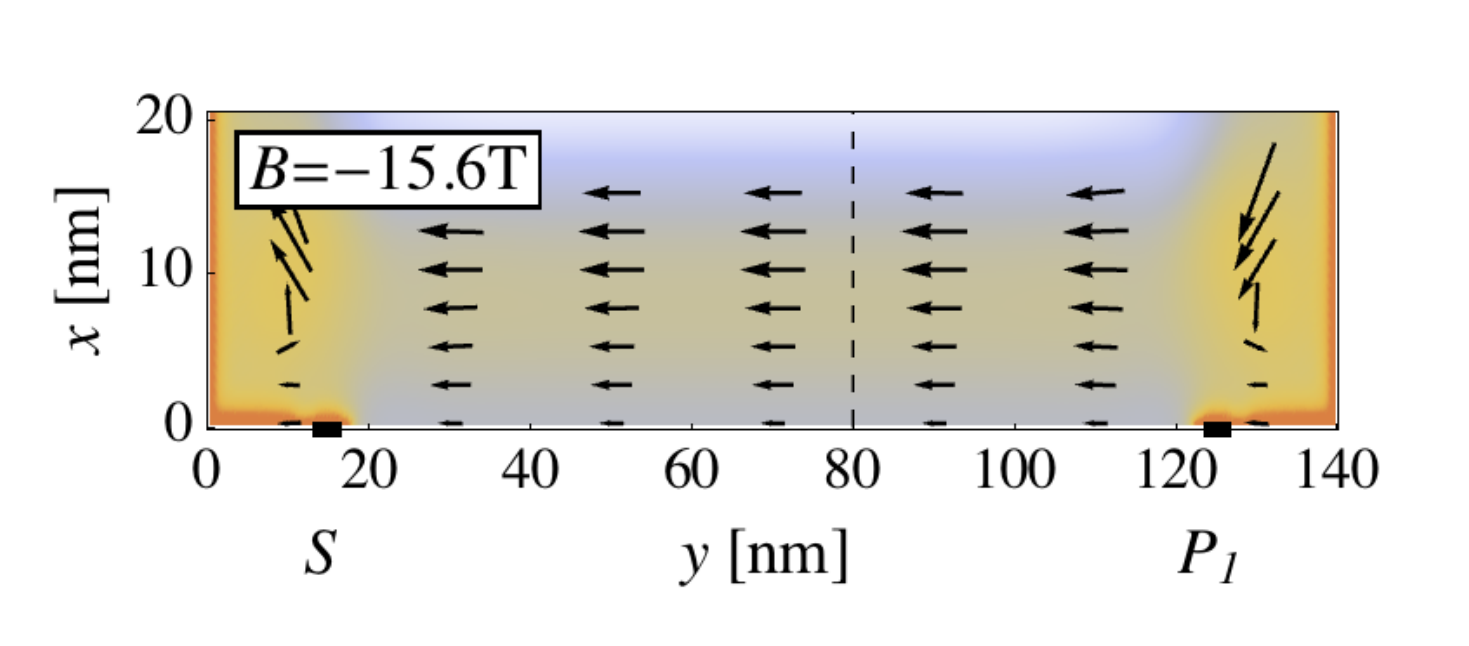} \includegraphics[scale=0.5]{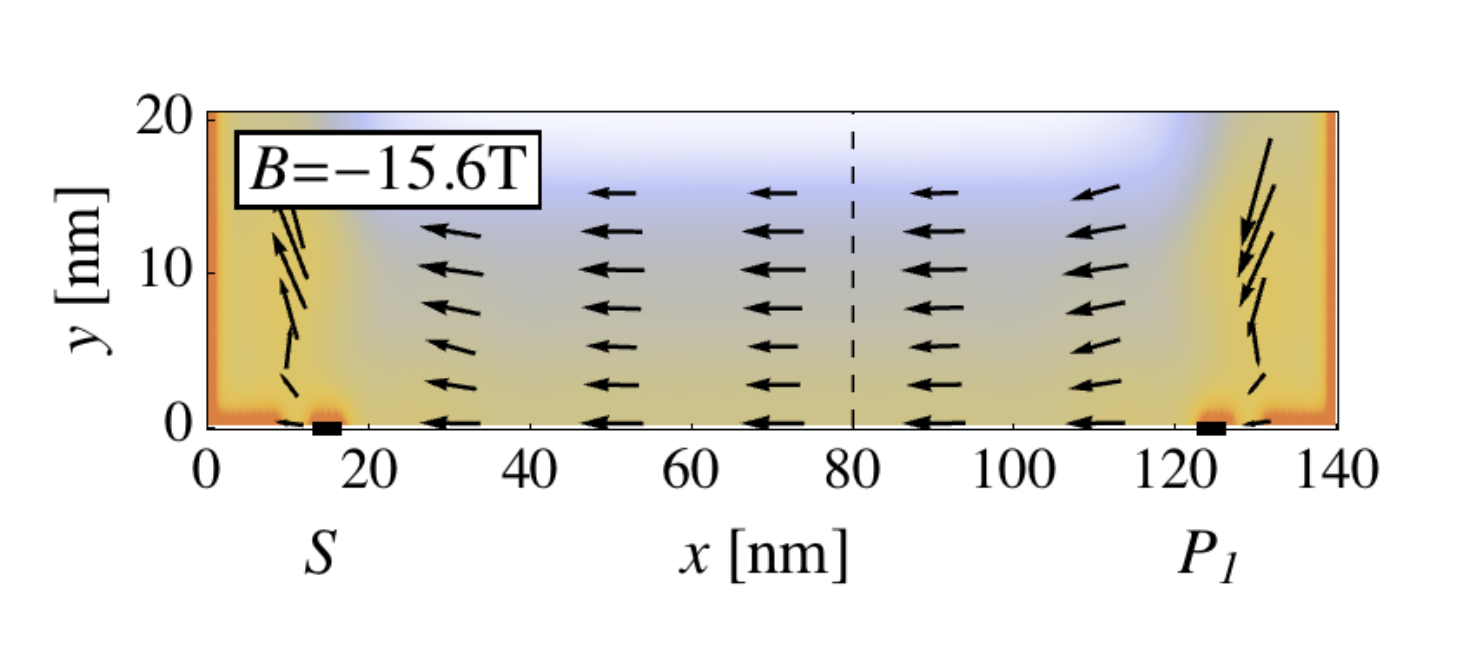}
  \caption{The edge channel transport of the quantum Hall effect can be observed clearly, when the direction
    of the magnetic field is reversed. A finite current flows at the armchair edge (right), whereas the
    current vanishes at the zigzag edge (left). This can also be seen in the transverse current through the
    dashed vertical line in \fig{9}. In the LDOS only a single broadened edge channel can be recognized,
    instead of spatially separated edge channels.}
  \label{fig:6}
\end{figure*}

\subsection{Cyclotron motion and the quantum Hall effect} \label{sec:cyc-mot-qhe}

The Hall resistance $R_{xy}$ as a function of the magnetic field $B$, calculated by means of the NEGF method
\eq{7}, is shown in \fig{4}. In the case of specular reflections at the boundary between $S$ and $P_1$ (blue
curve), the Hall resistance of both stripes shows at low magnetic field $0\un{T}< B < 10\un{T}$ a series of
equidistant peaks located approximately at
\begin{equation}
  \label{eq:23}
  B_n= \frac{2\mu}{e v_F L}\, n, \qquad n=1,2,3,\hdots,
\end{equation}
see the dashed vertical lines. At these magnetic fields a multiple $n$ of the cyclotron diameter
$2\Abs{\vec{p}}/eB$ equals the distance $L$ between injector and collector. Cyclotron orbits can be clearly
seen in \fig{5}, which shows the local current and the local density of states (LDOS) of electrons originating
from $S$ with energy $\mu$. Note that the shown local current and the LDOS have been averaged over the
honeycomb cells. These current flow paths can be measured by scanning tunneling microscopy \cite{Aidala2007}.

Extended quantum Hall plateaus in a strong magnetic field $B > 10 \un{T}$ can be observed, if the boundary in
between $S$ and $P_1$ is diffusive (red curve), or if the magnetic field is reversed and the current passes by
the other diffusive boundaries (see the end of section~\ref{sec:negf}). The current is carried through edge
channels along the boundaries, see \fig{6}. The quantum Hall effect in graphene can be understood easily by
the eigenenergy spectra shown in \fig{2}. The number of the occupied edge states at the Fermi energy equals
$2n+1$, where $n$ is the Landau level index. As every occupied edge state is a ballistic conductor, which
contributes with $2e^2/h$ to the total conductance, the Hall resistance reads
$ R_{xy}= \frac{h}{2e^2} \, \frac{1}{2n+1}, \, n=0,1,2, \hdots $.  This explains the quantum Hall staircase
observed in \fig{4}, which is one of the definitive fingerprints of a relativistic 2DEG \cite{Gusynin2005,
  Zhang2005, Novoselov2005, Novoselov2007, Jiang2007}, because it differs significantly from the
nonrelativistic case \cite{Datta1997}. In \fig{4} we can also observe that the transitions between the Hall
plateaus differ slightly in the two stripes. This can be explained by the shallow valleys in the band
structure at an armchair edge, which are not present at a zigzag edge or when the scattering at all boundaries
is diffusive.

\subsection{Anomalous resistance oscillations} \label{sec:ano-res-osc}

\begin{figure}[htb] 
  \centering
  \vspace*{1mm}
  \includegraphics[scale=0.82]{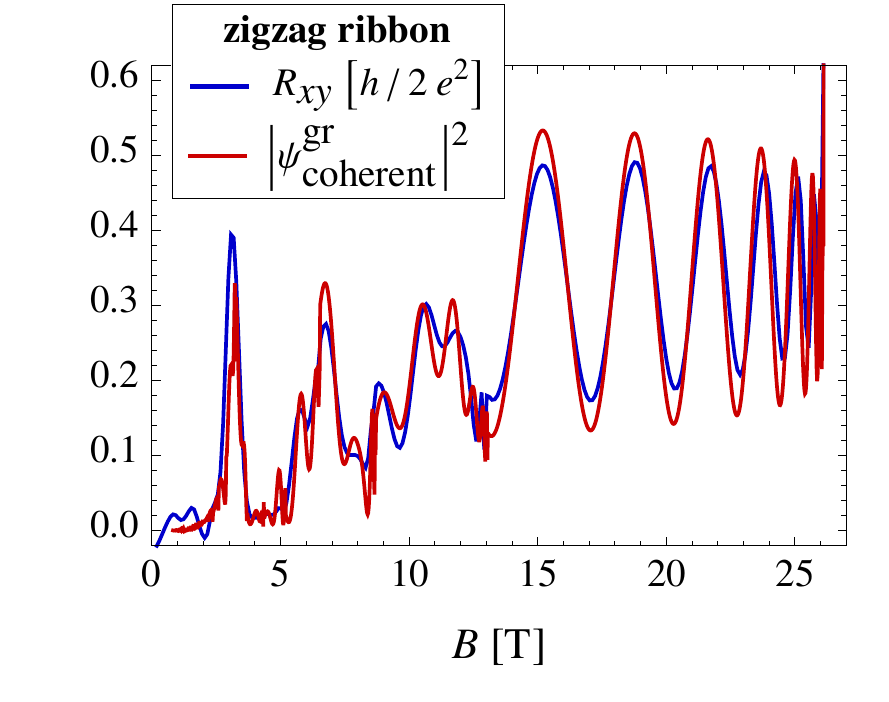}\\[3mm]
  \includegraphics[scale=0.82]{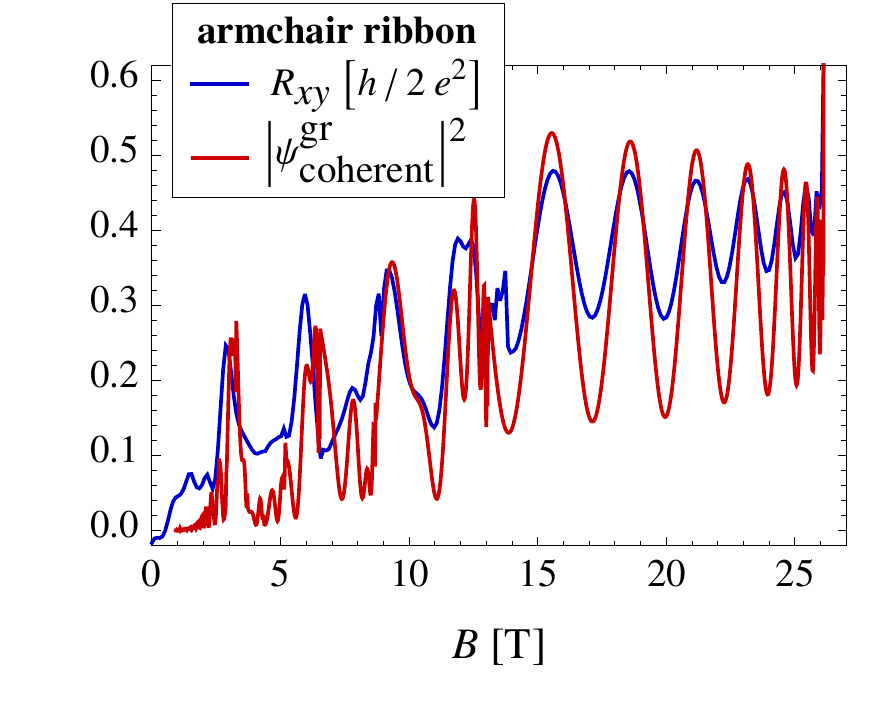}
  \caption{Normalized absolute square of the superimposed plane wave part of the occupied edge channels
    \eq{24}, (red curve) agrees well with the Green's function calculation of the Hall resistance (blue
    curve). All resistance oscillations can be understood by the interference of the edge channels. The
    anomalous oscillations are beatings, which appear when only some few edge channels are occupied.}
  \label{fig:7}
\end{figure}

In the case of specular scattering between $S$ and $P_1$ and in magnetic fields $B> 10 \un{T}$, we observe --
superimposed upon the quantum Hall plateaus -- anomalous resistance oscillations, which cannot be understood
by classical cyclotron motion. In particular, when only two Landau level are occupied ($16\un{T}<B<26\un{T}$),
the oscillations become very clear and regular. Their frequency increases rapidly whenever a Landau level is
pushed towards the Fermi energy and a transition between Hall plateaus appears (compare blue and red curves in
\fig{4}). Finally, the oscillations vanish completely, when only a single edge channel is occupied
($B>26\un{T}$), and the Hall plateau $R_{xy}=1$ appears (not shown in \fig{4}). These resistance oscillations
can be understood by means of the solution of the Dirac equation. In the zigzag stripe the edge states are
given by \eq{18} and \eq{19}. We superimpose the plane wave part of the occupied edge states
\begin{equation}
  \label{eq:24}
  \Abs{\psi_{\tx{gr}}}^2= \frac{ \Bigl\langle\Bigl|
  \sum_{i=1}^{n} \E^{\I\bigl(k_i +\frac{2\pi}{3\sqrt{3}}\bigr) L} 
  +\sum_{i=1}^{n+1} \E^{\I\bigl(q_i -\frac{2\pi}{3\sqrt{3}}\bigr) L}\Bigr|^2 \Bigr\rangle_{S,P_1}}{\lr{2n+1}^2},
\end{equation}
where $\langle \cdot \rangle_{S,P_1}$ means spatial averaging over the finite width of the injector and
collector contacts. The occupied edge states in the $\vec{K}$ valley are denoted by $k_i$ and the states in
the $\vec{K}'$ valley by $q_i$, see the red and blue dots in \fig{2}. The normalized absolute square of these
superimposed plane waves agrees almost perfectly with the NEGF calculation, see \fig{7} (top). Thus, all
focusing peaks can be understood by the interference of the plane wave part of the occupied edge states. The
anomalous resistance oscillations are beatings, which appear when only some few edge channels are
occupied. These beatings are very clear and regular if only two Landau levels are occupied. Their frequency
increases rapidly, whenever the highest occupied Landau level approaches the Fermi energy, because its
intersection point with the Fermi energy and thus, the corresponding $k_{\tx{max}}$ (or $q_{\tx{max}})$
increases strongly. The difference of $k_{\tx{max}}$ to the other, much smaller $k_n$ leads to a high
frequency beating. Finally, when only a single edge channel is occupied, the beating and thus, the
oscillations in the Hall resistance vanish. In the armchair stripe, the solution of the Dirac equation is more
complicated, see \eq{20}, \eq{21}, and \eq{22}, because the valleys are intermixed. We found best agreement to
our Green's function calculations, see \fig{7} (bottom), if we use also for the armchair stripe \eq{24}, where
the $k_i$ and $q_i$ denote the two sets of solutions.

In a magnetic field $B<16\un{T}$, the simplified model shows smaller highly oscillating peaks, which
are due to the interference of numerous plane waves. These highly oscillating peaks are more
pronounced in the armchair stripe, where additional interference between the $\vec{K}$ and
$\vec{K}'$ valley takes place, which is not present in the zigzag stripe, compare \eq{21} and
\eq{19}. The highly oscillating peaks are not present in the NEGF calculations due to the diffusive
boundaries. Probably they neither appear in the experiment due to the presence of decoherence. In a
stronger magnetic field, the highly oscillating peaks disappear and the simplified model agrees very
well with the NEGF calculations, because the superposition of only few eigenstates (see \fig{2})
leads to beatings. The oscillation are almost independent from the edge geometry, apart from slight
differences in their frequency and phase.

\begin{figure*}[htb]
  \begin{minipage}[c]{0.49\linewidth}
    \includegraphics[scale=0.8]{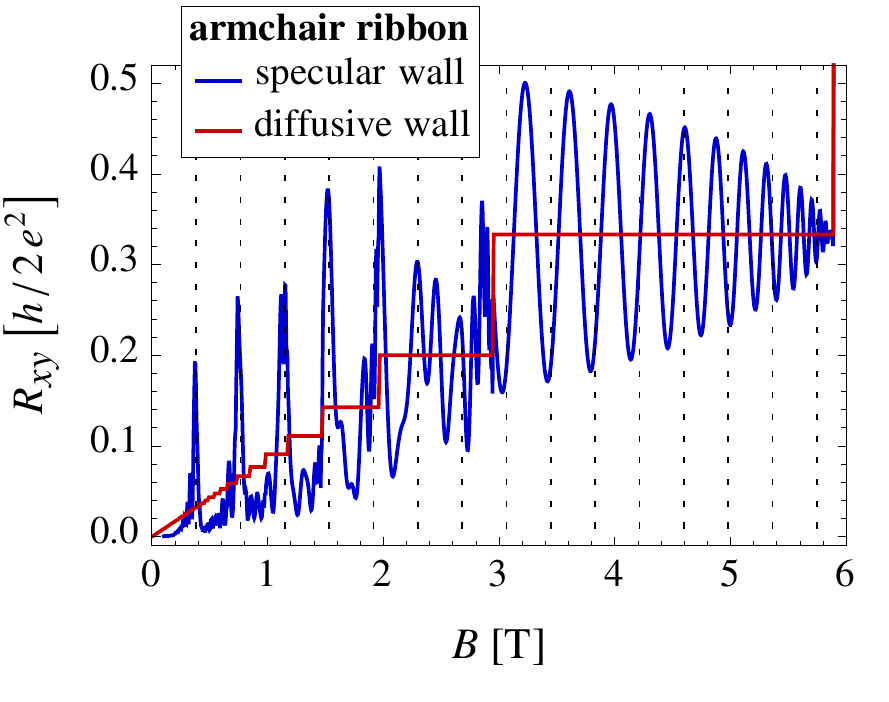}
  \end{minipage}
  \begin{minipage}[c]{0.48\linewidth}
    \hspace*{1.5mm}\includegraphics[scale=0.8]{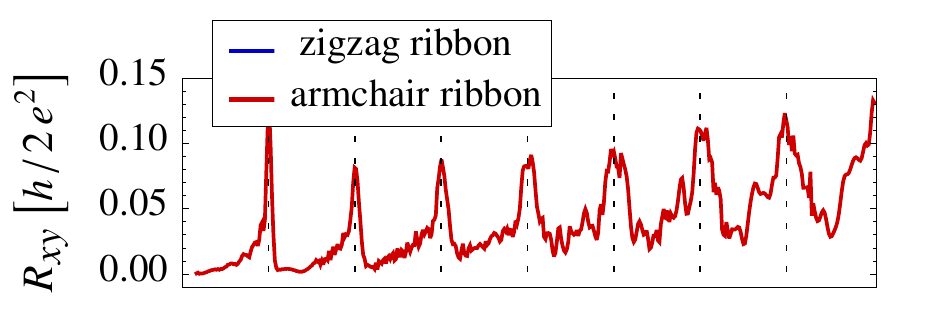}\\[-4mm]
    \hspace*{1.5mm}\includegraphics[scale=0.8]{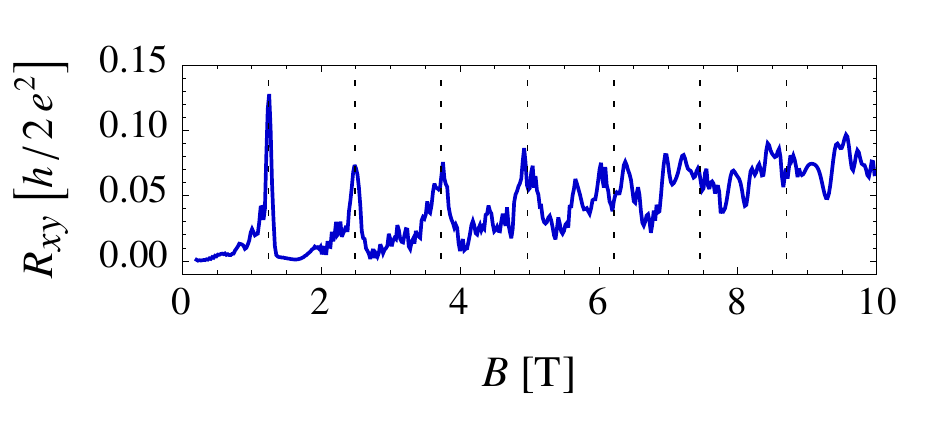}
  \end{minipage}
  \caption{Hall resistance of larger graphene stripes calculated by means of \eq{24}. Contacts with a width of
    $20 \un{nm}$ are attached at a distance of $450 \un{nm}$. Right: When the Fermi energy is set to
    $\mu= 260 \un{meV}$, corresponding to a carrier density of $n_{\tx{gr}}= 6.0 \cdot 10^{12} \un{cm^{-2}}$,
    only classical equidistant focusing peaks can be observed. As reported in \cite{Rakyta2010}, the classical
    focusing peaks of higher order ($n>4$) are clearly visible at armchair edges (top) but are suppressed at
    zigzag edges (bottom). Left: When Fermi energy $\mu= 80 \un{meV}$ and carrier density
    $n_{\tx{gr}}= 5.7 \cdot 10^{11} \un{cm^{-2}}$ are lowered, we find classical equidistant focusing peaks
    followed by anomalous oscillations.}
  \label{fig:8}
\end{figure*}

The beatings, which appear in the case of only two occupied Landau levels, can be used to determine precisely
the distance between the injector $S$ and collector $P_1$. In \fig{7}, the almost perfect match of the
positions of all extrema in the range $13 \un{T} < B <26 \un{T}$ is obtained only, if $L=110 \un{nm}$ is
chosen in \eq{24} for the distance between $S$ and $P_1$. In order to explain, why in armchair stripes the
classical focusing peaks deviate slightly from their expected positions, see \fig{4} (bottom), we could assume
hypothetically a slightly larger distance $L=120 \un{nm}$ between injector and collector. In this case,
the classical focusing peaks would appear exactly at the expected positions, but the beatings would absolutely
not fit to \eq{24}. Also finite size effects can be ruled out as these deviations are not present in zigzag
stripes of the same size. One reason for the shift of the classical focusing peaks could be the distinct edge
current observed only at armchair edges or edge dependent scattering \cite{Petrovic2015}.

Although the charge carriers in graphene behave as relativistic massless fermions, the studied stripes show
properties similar to a nonrelativistic 2DEG \cite{Stegmann2013}: Classical focusing peaks in weak magnetic
fields, followed by anomalous resistance oscillations when the magnetic field strength is increased. In both
systems the resistance oscillations can be explained by the interference of the plane wave part of the
occupied edge states. However, in graphene the linear dispersion, the valley degeneracy (symmetry points
$\vec{K}$ and $\vec{K}'$ in momentum space) as well as the non-trivial edge geometry add subtle but important
new aspects. In this way, at first sight the local current flow looks similar in both systems (cyclotron
orbits, edge channels), compare \fig{5} with figure~2 in \cite{Stegmann2013}. However, the boundary geometry
has a distinct effect on the local current flow, see \fig{6}, which will be discussed in
section~\ref{sec:edg-cur-flw}.

\subsection{Experimental observability} \label{sec:exp-obs}

Due to computational limitations, the studied stripes are relatively small ($L=110 \un{nm}$) and the
considered magnetic fields are quite strong ($B_{\tx{max}}=30 \un{T}$). In these strong fields, also the
Zeeman spin splitting of the Landau levels can be relevant \cite{Zhang2006, Jiang2007, Gusynin2008} but we do
not expect that the spin splitting changes qualitatively our findings. Although it is technically possible to
realize such system parameters, this is not essential to observe our findings in an experiment. The important
factor in an experiment is the maximal number of resolvable focusing peaks $n_{\tx{max}}$, which is limited
due to decoherence and partial diffusive scattering at the boundary. In order to observe anomalous resistance
oscillations due to the interference of some few edge channels, the distance $L$ between injector and
collector as well the Fermi energy $\mu$ have to be tuned in such a way that the maximal number of possible
specular reflections fulfills the rule of thumb
\begin{equation}
  \label{eq:25}
  n_{\tx{max}} \sim  \frac{1}{6} \, \frac{L}{a}\,\frac{\mu}{t}, 
\end{equation}
which can be derived easily by \eq{14} and \eq{23}. Of course, mean free path and phase coherence length also
have to be comparable with $ L $. To our knowledge in most focusing experiments such system parameters have
been used that the regime of coherent electron focusing and the quantum Hall effect are well separated, see
e.g. Figure~10 in \cite{Houten1989}. However, signs of the anomalous oscillations can be observed in different
geometries \cite{Ford1990, Ford1991}.

\begin{figure}[htb] 
  \centering
  \includegraphics[scale=0.82]{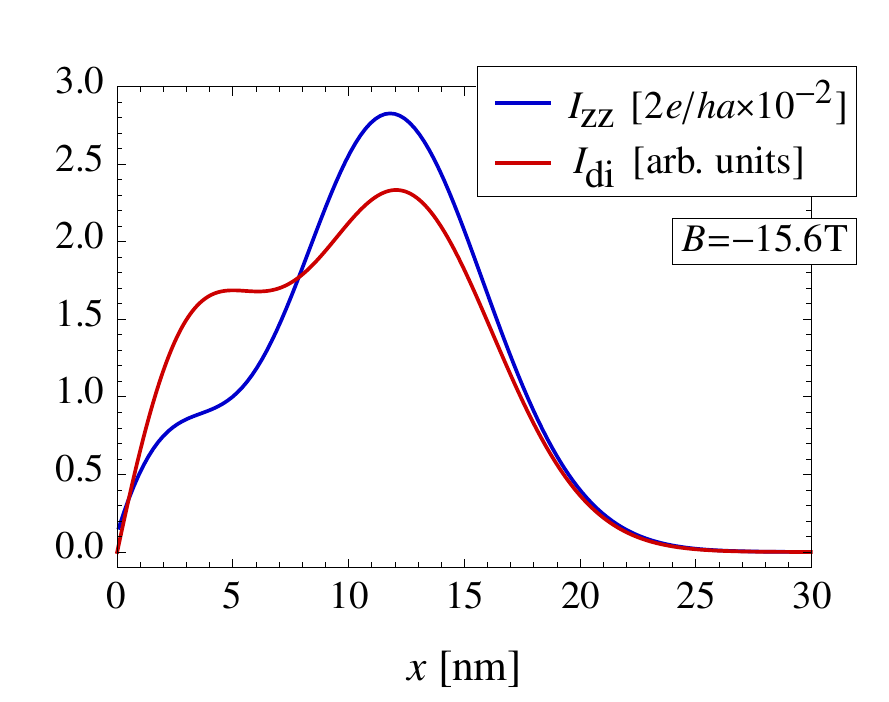}\\
  \includegraphics[scale=0.82]{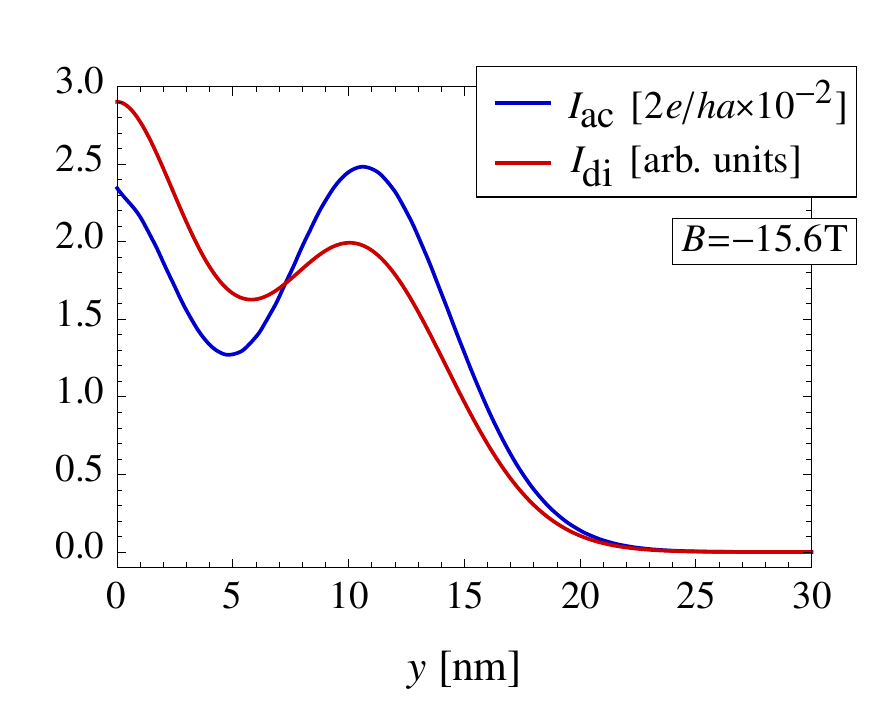}
  \caption{Transverse current (blue curves) through the dashed lines in \fig{6}. A finite current flows on the
    armchair edge (bottom) but the current vanishes on the zigzag edge (top). The current, calculated by the
    eigenstates of the Dirac equation (red curve), agrees with the NEGF calculation and allows to attribute
    the different edge currents to the different boundary conditions of the stripes. We can identify two
    spatially separated edge channels, which equals the number of occupied Landau levels (with $E\geq 0$). Due
    to the boundary conditions, the edge channels are more densely packed in the zigzag stripe.}
  \label{fig:9}
\end{figure}

\begin{figure}[htb]
  \centering
  \vspace*{-0.2mm}
  \includegraphics[scale=0.514]{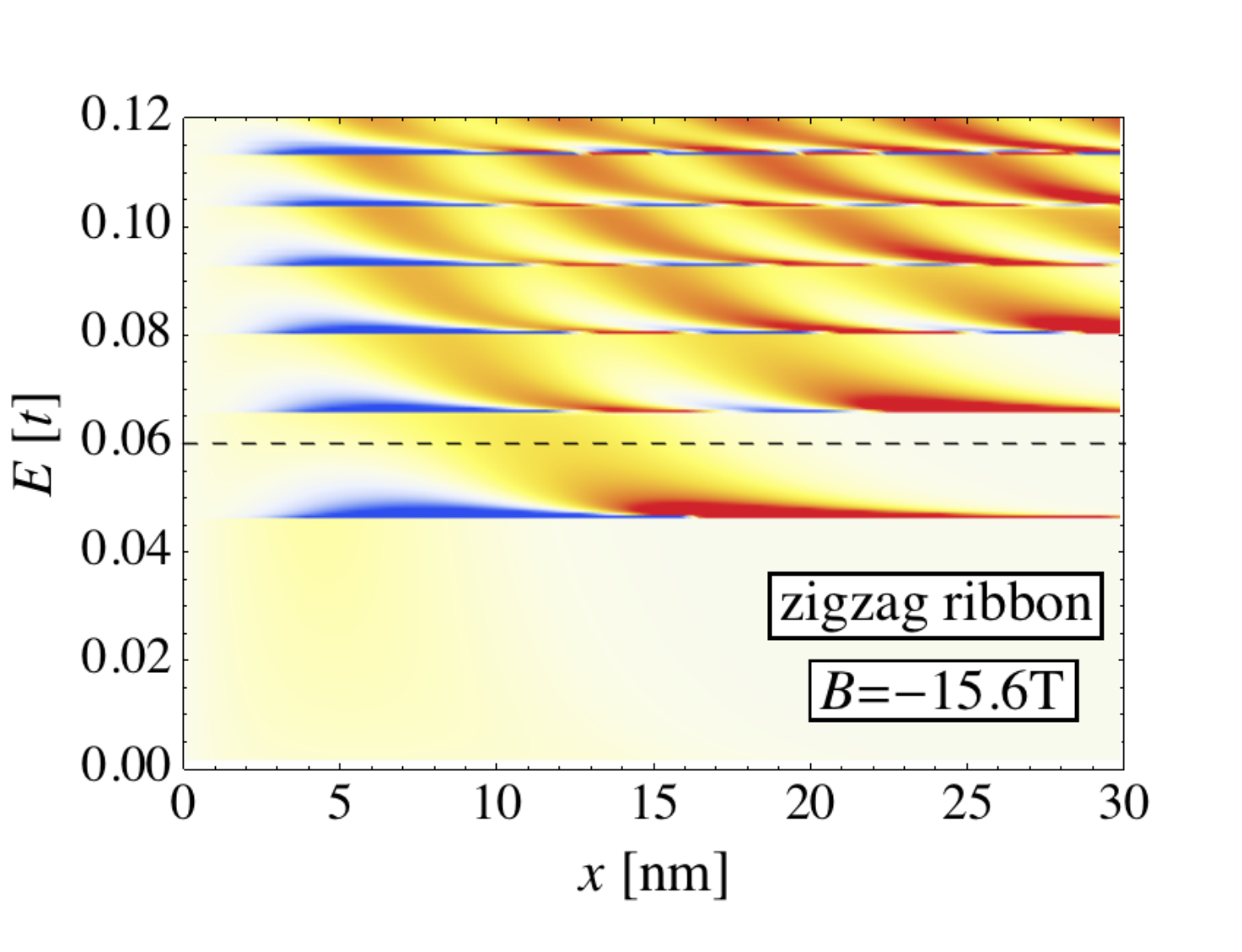}\\[1.3mm]
  \includegraphics[scale=0.514]{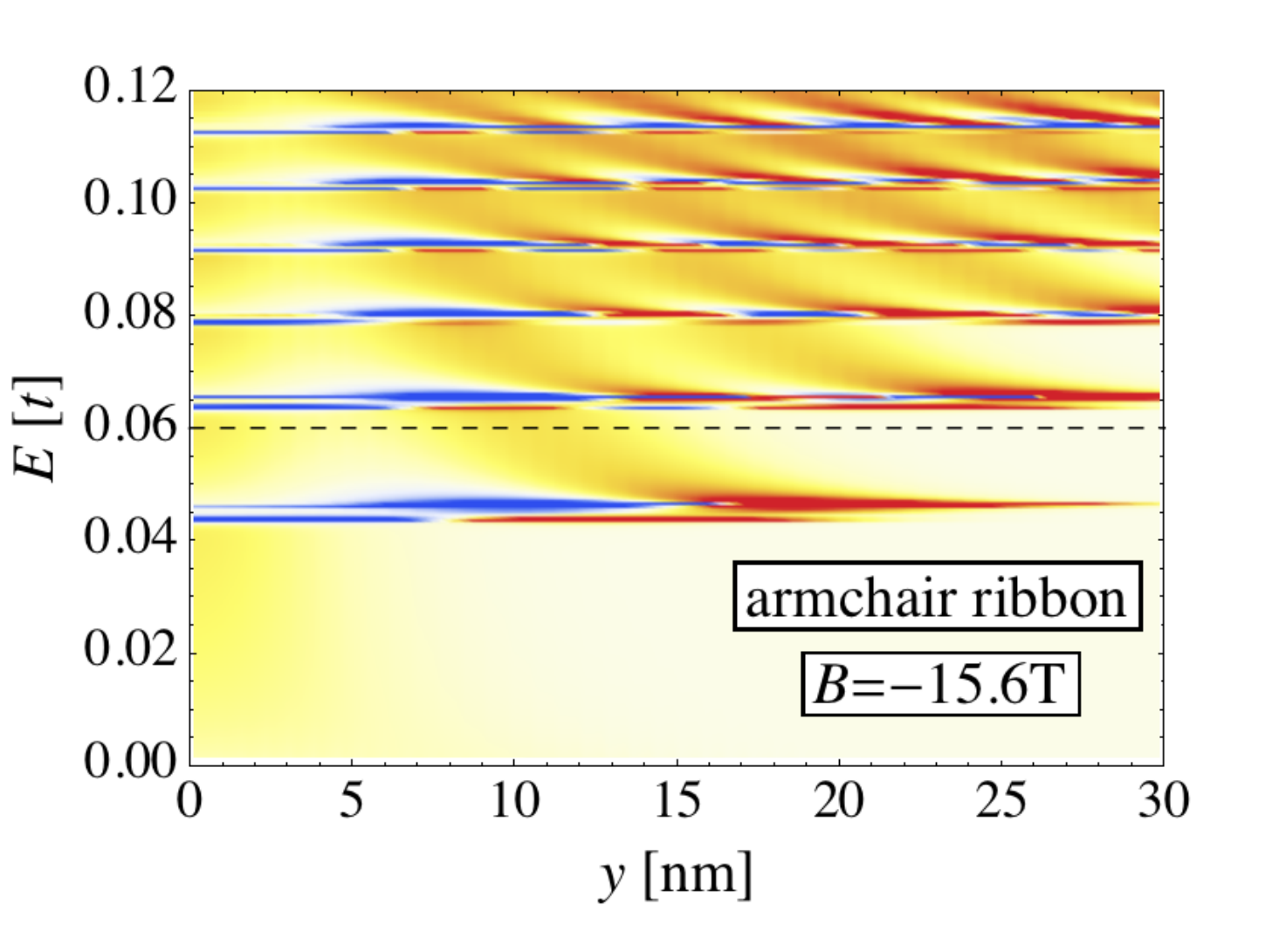}
  \caption{Energy resolved transverse current through the dashed lines in \fig{6}. Warm colors indicate a
    current from $P_1$ to $S$ while cold colors correspond to a current in the opposite direction. As shown by
    the shading close to the edges of the stripes (i.e. close to $x=0$ and $y=0$, respectively), a finite
    current flows on the armchair edge (bottom), which is not present on the zigzag edge (top). The number of
    spatially separated edge channels equals the number of occupied Landau levels (with $ E \geq 0$), although
    the two edge channels closest to a zigzag edge are hardly distinguishable. Surprisingly, close to the
    Landau levels regions of counterpropagating current can be observed (blue regions). However, the total
    (integrated) current is quantized and does not change its sign.}
  \label{fig:10}
\end{figure}

Because of the excellent agreement of the simplified model \eq{24} and the Green's function calculations, see
\fig{7}, we can use this simplified model to study larger stripes, for which NEGF calculations are
demanding. We consider stripes at which $20 \un{nm}$ wide contacts are attached at a distance of
$450\un{nm}$. This is approximately the same geometry used in the recent focusing experiment in graphene
\cite{Taychatanapat2013} as well as in a theoretical study \cite{Rakyta2010}. When the Fermi energy is set to
$\mu= 260 \un{meV}$ corresponding to a carrier density of $n_{\tx{gr}}= 6.0 \cdot 10^{12} \un{cm^{-2}}$, the
system is in the regime of classical equidistant focusing peaks ($n_{\tx{max}} \sim 49 $), see \fig{8}
(right). In agreement with results reported by Rakyta et al. \cite{Rakyta2010}, the focusing peaks of higher
order ($n>4$) are clearly visible at armchair edges but are suppressed at zigzag edges. When Fermi energy
$\mu= 80 \un{meV}$ and carrier density $n_{\tx{gr}}= 5.7 \cdot 10^{11} \un{cm^{-2}}$ are lowered, we bridge
the regime of coherent electron focusing and the quantum Hall regime ($n_{\tx{max}}\sim 15$), see \fig{8}
(left). The Hall resistance starts with equidistant classical peaks, but anomalous oscillations follow when
the strength of the magnetic field is increased. This gives us confidence that the predicted resistance
oscillations can be observed experimentally.

\subsection{Edge current flow} \label{sec:edg-cur-flw}

In figures~\ref{fig:5} and \ref{fig:6} we observe that a finite current flows on the armchair edge, whereas
the current vanishes on the zigzag edge. This can be seen clearly in \fig{9} (blue curve), which shows the
transverse current through the dashed vertical lines in \fig{6}.  It can be understood, if we calculate the
transverse current by means of the eigenstates of the Dirac equation \cite{MunozRojas2006, Katsnelson2006,
  Wang2011, Katsnelson2012}
\begin{equation}
  \label{eq:26}
  I_{\tx{di}}(\vec{r}) \propto \sum_{i=1}^{2n+1} \psi_{A,k_i} \psi_{B,k_i} \propto \sum_{i=1}^{2n+1} c_i \,
  \mc{D}_{\nu,k_i} \mc{D}_{\nu-1,k_i},
\end{equation}
where $c_i $ is a normalization constant and the sum is over the occupied edge states, see the dots in
\fig{2}. At zigzag edges the parabolic cylinder functions have to be zero, see \eq{19}, which results in zero
edge current. At armchair edges, the sum of the parabolic cylinder functions has to be zero, see \eq{21},
which allows for a finite edge current. The transverse current calculated by means of \eq{26} agrees well with
the Green's functions calculations, see the red curves in \fig{9}. In the transverse current at
$B=-15.6 \un{T}$, we can identify two spatially separated edge channels. Thus, the number of spatially
separated edge channels in the local current, averaged over the honeycomb cells, equals the number of occupied
Landau levels (with energy $E\geq 0$), compare with \fig{2}. The lifting of their degeneracy at the edge is
not resolved in the local current. Due to the boundary conditions, in the zigzag stripe the two edge channels
are more densely packed and harder to separate than in the armchair stripe. Note that the total edge current
is approximately independent from the edge geometry. The energy resolved transverse current in \fig{10}
confirms these findings. Surprisingly, it also shows counterpropagating currents close to the Landau levels,
see the blue shaded regions, in which the current flows in the opposite direction as in the red shaded
regions. However, note that the total (integrated) current is quantized and does not change its sign. The
counterpropagating currents are also found when the NEGF method is applied to a nonrelativistic 2DEG. At this
point, their origin is not understood, but they are also observed by Wang et al. \cite{Wang2011} using the
eigenstates of the Dirac equation. Also the dependency of the current on the edge geometry is reported in
their work. Beyond that, we show in \fig{5} that a distinct armchair edge current appears also in the regime
of coherent electron focusing. This distinct armchair edge current in focusing experiments could be measured
experimentally by means of an additional voltage probe placed on the stripe's edge or by contacting edge
channels individually, as in \cite{Wuertz2002, Deviatov2006}.

\subsection{Local density of states} \label{sec:loc-dens-stat}

\begin{figure}[htb]
  \centering
  \includegraphics[scale=0.8]{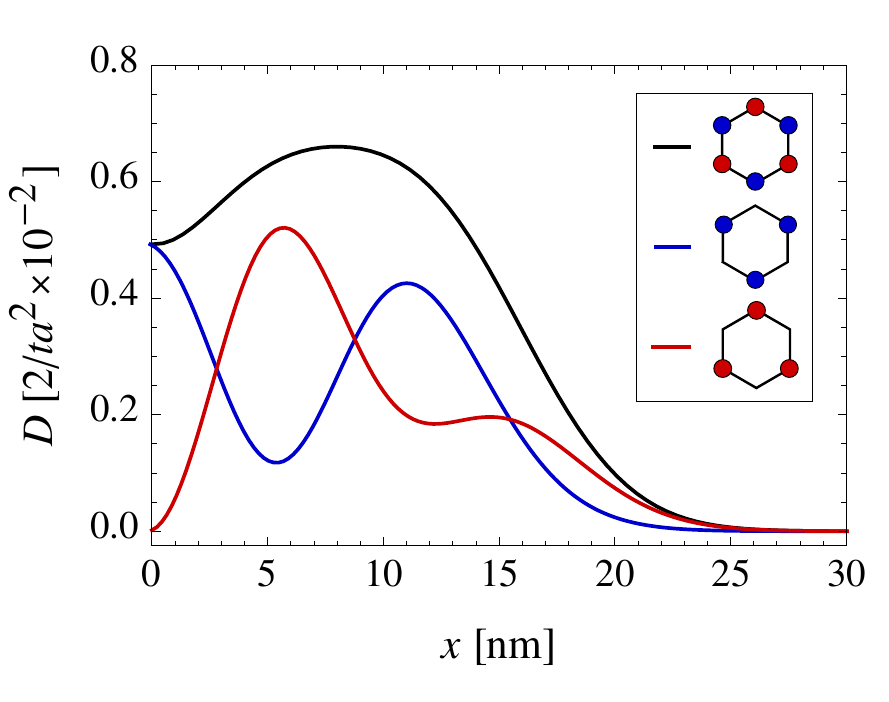}
  \includegraphics[scale=0.8]{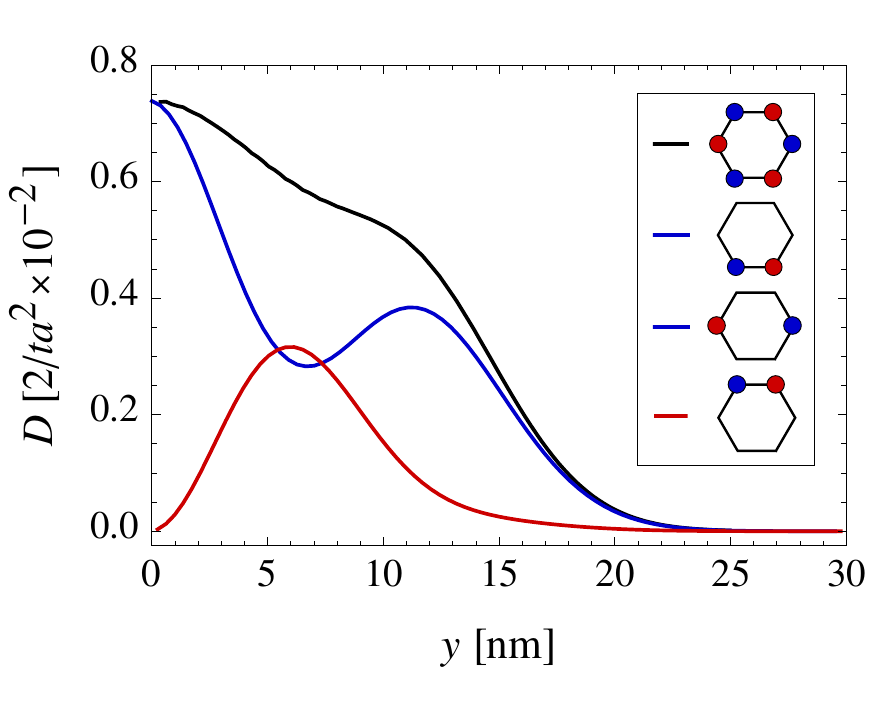}
  \caption{LDOS of the zigzag stripe (top) and the armchair stripe (bottom) along the dashed vertical lines in
    \fig{6}. The LDOS averaged over the six carbon atoms of the honeycomb cells (black curve) shows a single
    broadened edge channel. Individual edge channels become visible, when only a subset of the atoms is taken
    into account, see marked atoms in the legend. Note that in the armchair stripe two subsets give
    numerically identical results (blue curve).}
  \label{fig:11}
\end{figure}

The local density of states (LDOS) in \fig{6}, averaged over the six carbon atoms of the honeycomb cells,
shows only a single broadened edge channel. This can be seen clearly in \fig{11} (black curves), which gives
the LDOS along the dashed vertical line in \fig{6}. In order to make individual edge channels visible in the
LDOS, we have to select only a subset of the carbon atoms, see the blue and red curves for which only the
atoms marked in the inset are taken into account. Note that in the armchair stripe two subsets give
numerically identical results, see the blue curve. When every carbon atom is considered individually, the LDOS
oscillates rapidly between the blue and red curves in \fig{11}. These oscillations have been reported in
theoretical studies \cite{Brey2006, Brey2006_2, MunozRojas2006, Zarbo2007}, but to our knowledge an
experimental confirmation is missing. The energy resolved LDOS, calculated numerically by means of the NEGF
method, is depicted in \fig{12}. Far from the edge the discrete Landau levels can be observed clearly. If the
LDOS is averaged over the honeycomb cells (left column), the bending of the energy bands can hardly be
discerned. It becomes more visible, if only a subset of the carbon atoms is taken into account (middle and
right column). In this way, we can observe how in the zigzag stripe (top row) the zeroth Landau level at $E=0$
splits into a dispersive edge state on the sublattice B (right) and a non-dispersive surface state on the
sublattice A (middle). This surface state is not present in the armchair stripe. Similar results can also be
obtained by means of the eigenstates of the Dirac equation, see \cite{Abanin2007}. Anyway, in the experiment
it is not possible to select a subset of the carbon atoms. Thus, the measured LDOS looks similar to the
figures in the left column, see \cite{Li2013}.

\section{Conclusions} \label{sec:conclusions}

\begin{figure*}[htb]
  \centering
  \includegraphics[scale=0.5]{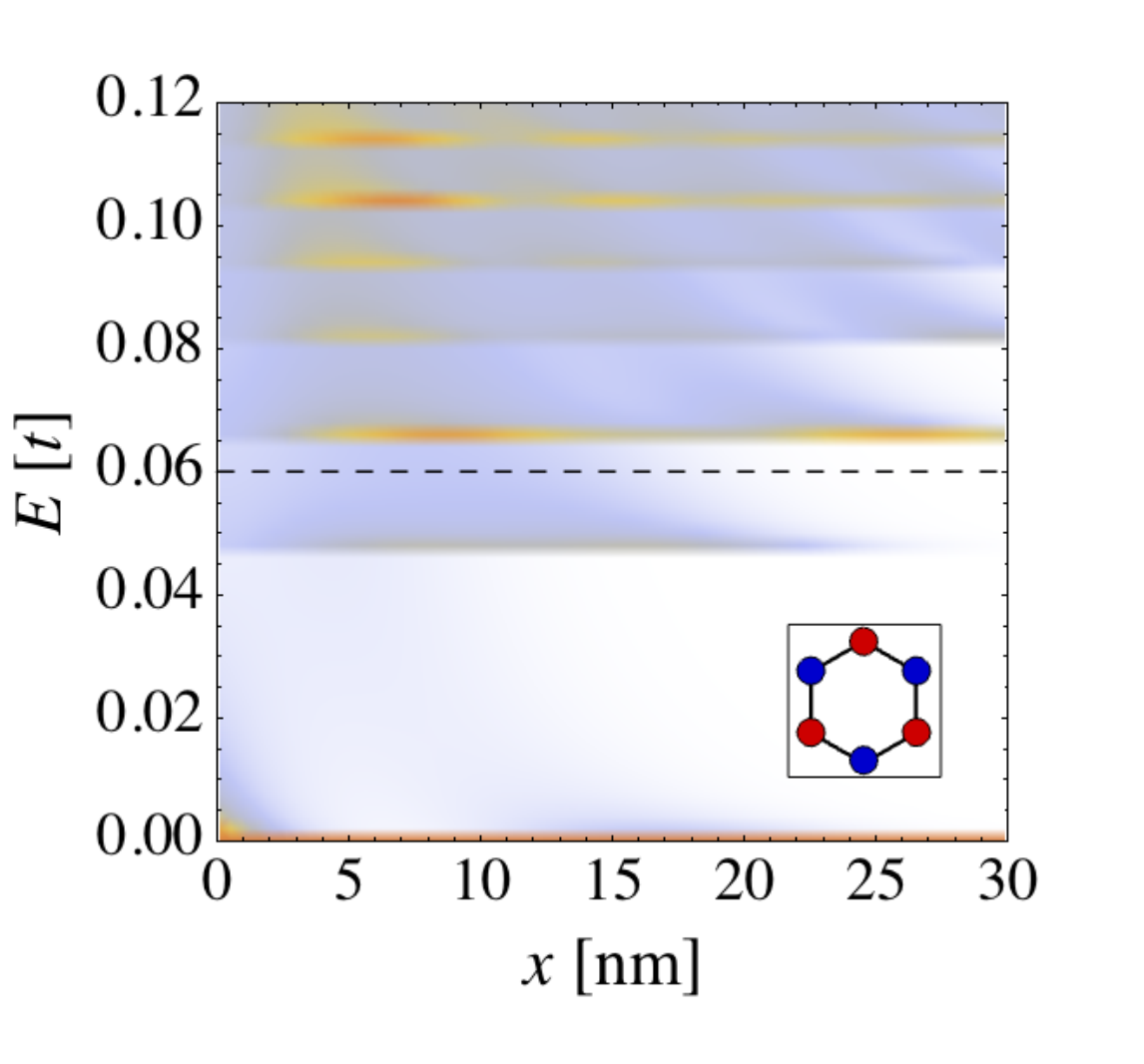}
  \hspace*{-4mm}
  \includegraphics[scale=0.5]{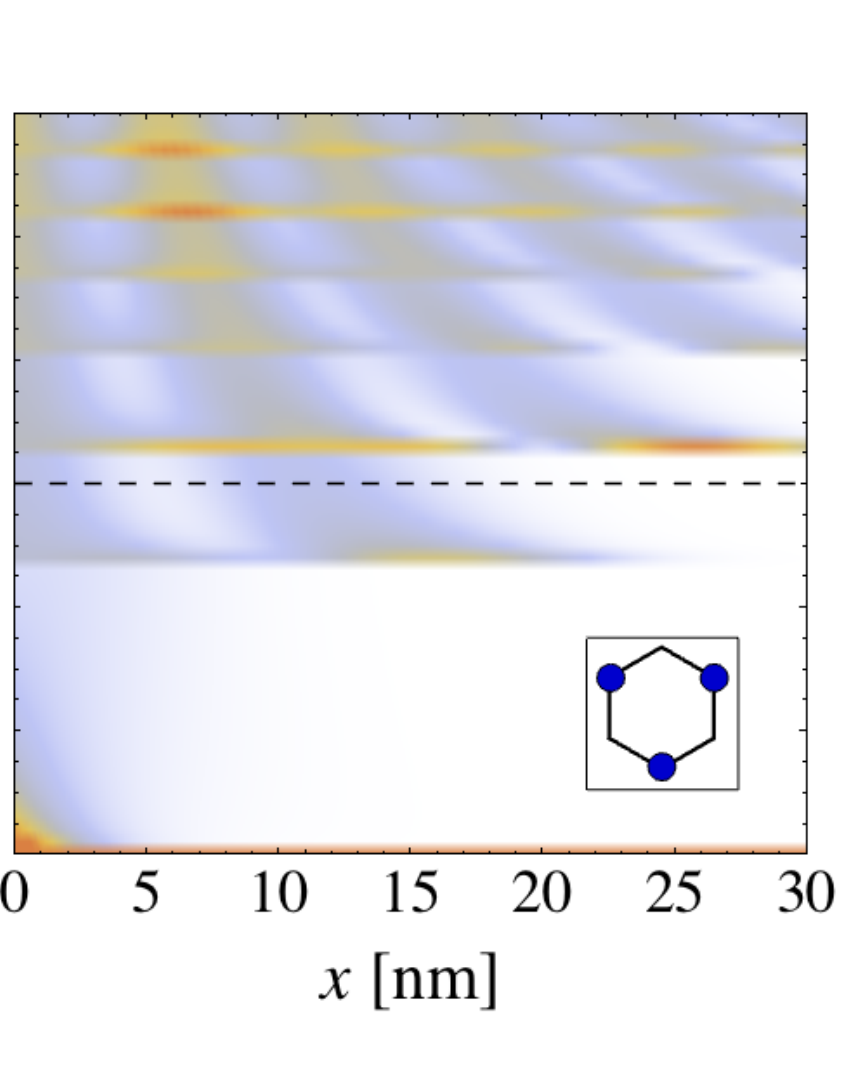}
  \hspace*{1.2mm}
  \includegraphics[scale=0.5]{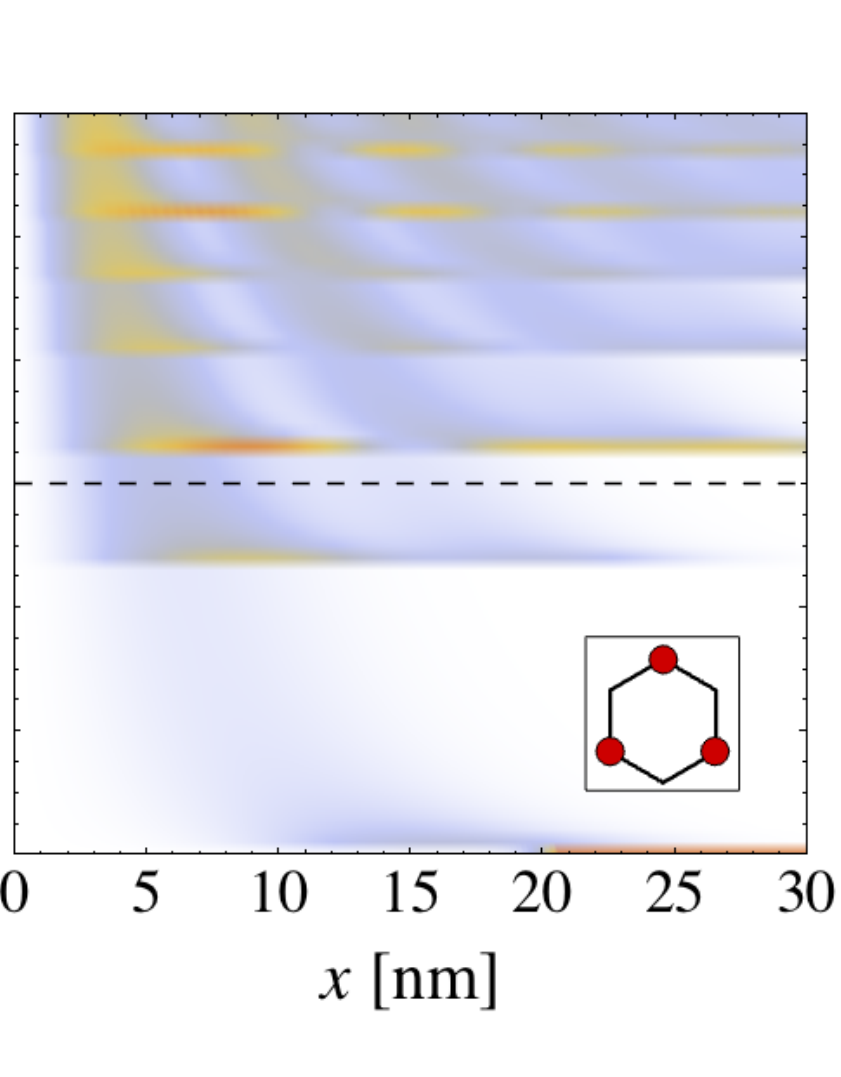}\\[-1.5mm]
  \includegraphics[scale=0.5]{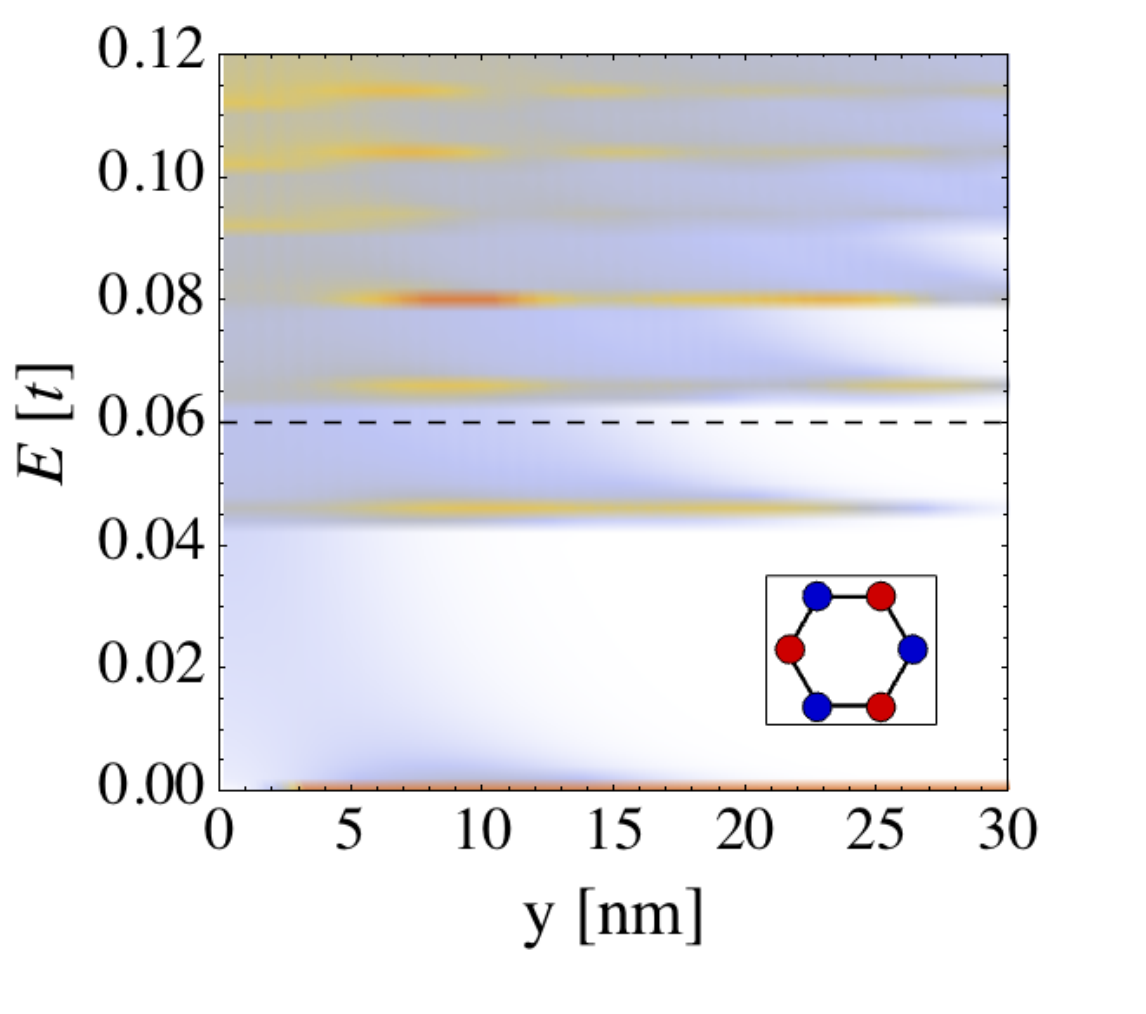}
  \hspace*{-4mm}
  \includegraphics[scale=0.5]{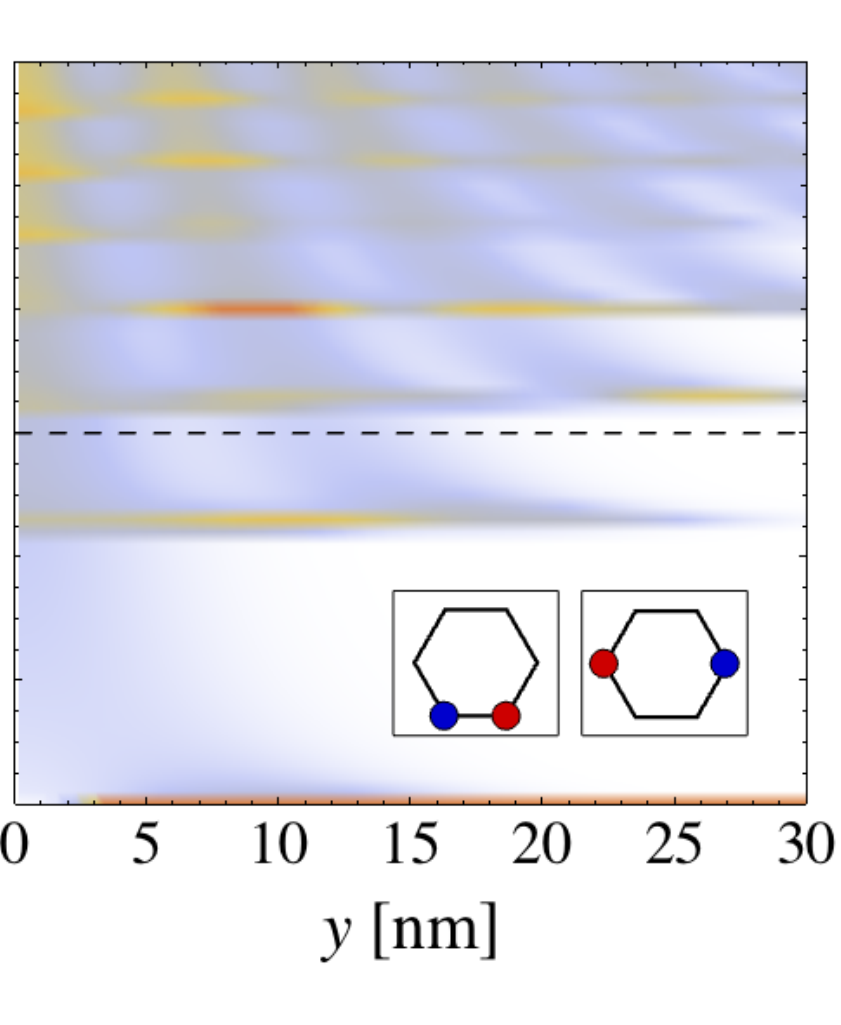}
  \hspace*{1.2mm}
  \includegraphics[scale=0.5]{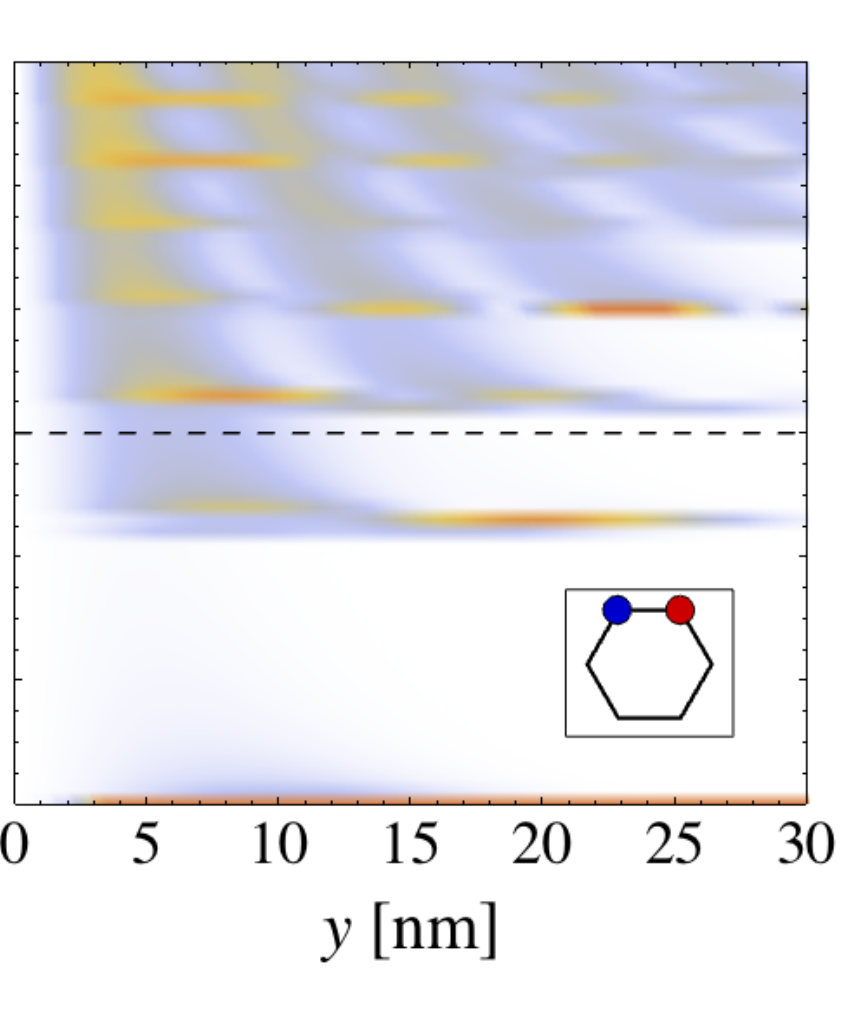}
  \caption{Energy resolved LDOS in the studied zigzag stripe (top row) and armchair stripe (bottom row). In
    the left column the LDOS has been averaged over the six carbon atoms of the honeycomb cells. Landau levels
    can be observed far from the edges. However, the bending of the energy bands in vicinity of the edge is
    seen more clearly in the figures of the middle and right column, where the LDOS is averaged only over the
    subset of atoms shown in the inset. In the zigzag stripe it can be seen how the zeroth Landau level (at
    $E=0$) splits into an edge state (top, right) and a non-dispersive surface state (top, middle), when the
    edge is approached. This surface state is not present in the armchair stripe.}
  \label{fig:12}
\end{figure*}

In this paper, we have studied theoretically magnetotransport in graphene stripes. In these stripes electrons
are injected at one point of the boundary and focused by a perpendicular magnetic field onto another point of
that boundary, see \fig{1}. We have calculated by the NEGF method the generalized Hall resistance as a
function of the magnetic field, see \fig{4}. In weak fields equidistant focusing peaks appear, which
correspond to classical cyclotron orbits \eq{23}, see \fig{5}. When the magnetic field is increased, anomalous
resistance oscillations are observed, which cannot be explained by classical cyclotron motion.

By means of a simplified model, we have shown that all calculated resistance oscillations can be understood by
the interference of the plane wave part of the occupied edge channels, see \fig{7}. The anomalous resistance
oscillations are beatings, which appear when only some few edge channels are occupied and only some few plane
waves are superimposed. Thus, the oscillations are very clear and distinct, if only two Landau levels are
occupied. The frequency of the resistance oscillations increases rapidly, when the magnetic field is increased
and a Landau level is depleted, because the momentum of the corresponding plane wave (and hence, its
frequency) is also increasing rapidly, see \fig{2}. Due to computational limitations, the studied graphene
stripes have been relatively small and the magnetic field has been relatively strong. However, due to the good
agreement of the simplified model with the NEGF calculations, we have used this model to show that our
findings are expected to appear also in larger stripes at lower magnetic fields. As the resistance
oscillations, classical focusing peaks as well as the beatings, are due to the interference of the edge
channels, we have also given a rule of thumb \eq{25} for the required number of specular reflections.

Studying the effect of the edge shape of the graphene stripes on the magnetotransport, we found that a finite
current flows on the armchair edge, whereas the current vanishes on the zigzag edge, see figures~\ref{fig:6}
and \ref{fig:9}. By means of the simplified model, the different edge currents can be traced back to the fact
that at an armchair edge carbon atoms of both sublattices appear, while at a zigzag edge only atoms of one
sublattice are present, see the inset of \fig{4}. We have also shown in figures~\ref{fig:9} and \ref{fig:10}
that the number of spatially separated edge channels in the local current equals the number of occupied Landau
levels. The discrete Landau levels can be seen clearly in the LDOS in \fig{12}. However, the bending of the
Landau levels in vicinity of the edge as well as spatially separated edge channels can be hardly recognized,
if the LDOS is averaged over the six carbon atoms of the honeycomb cells. They can be made visible, if the
LDOS is averaged only over a subset of the carbon atoms.

\begin{acknowledgement}
  We thank Dietrich E. Wolf for many inspiring discussions and helpful remarks. T. S. acknowledges a
  postdoctoral fellowship from DGAPA-UNAM and financial support from CONACyT research grant 154586 and
  PAPIIT-DGAPA-UNAM research grants IG101113 and IN114014.
\end{acknowledgement}

\bibliography{./Stegmann2015}

\end{document}